\documentstyle[aps,pre,epsfig,psfig,eqsecnum]{revtex}

\def\eps{\varepsilon}
\def\k{{\bf k}}

\def\const{{\rm const\,}}
\def\Sym{{\cal S}}

\begin{document}

\title{Anomalous scaling of a passive scalar advected by the
Navier--Stokes velocity field: Two-loop approximation}

\author{L.\,Ts.\,Adzhemyan$^{1}$,  N.\,V.\,Antonov$^{1}$,
J.\,Honkonen,$^{2}$ and T.\,L.\,Kim$^{1}$}

\address{$^{1}$ Department of Theoretical Physics, St.~Petersburg University,
Uljanovskaya 1, St.~Petersburg---Petrodvorez, 198504, Russia \\
$^{2}$ Theory~Division, Department~of~Physical Sciences,
P.O.~Box~64, FIN-00014 University~of~Helsinki, Finland}

\draft


\maketitle

\begin{abstract}
The field theoretic renormalization group and operator product
expansion are applied to the model of a passive scalar quantity
advected by a non-Gaussian velocity field with finite correlation
time. The velocity is governed by the Navier--Stokes equation,
subject to an external random stirring force with the correlation
function $\propto \delta(t-t') k^{4-d-2\varepsilon}$. It is shown
that the scalar field is intermittent already for small
$\varepsilon$, its structure functions display anomalous scaling
behavior, and the corresponding exponents can be systematically
calculated as series in $\varepsilon$. The practical calculation
is accomplished to order $\varepsilon^{2}$ (two-loop
approximation), including anisotropic sectors. Like for the
well-known Kraichnan's rapid-change model, the anomalous scaling
results from the existence in the model of composite fields
(operators) with negative scaling dimensions, identified with the
anomalous exponents. Thus the mechanism of the origin of anomalous
scaling appears similar for the Gaussian model with zero
correlation time and non-Gaussian model with finite correlation
time. It should be emphasized that, in contrast to Gaussian
velocity ensembles with finite correlation time, the model and the
perturbation theory discussed here are manifestly Galilean
covariant. The relevance of these results for the real passive
advection, comparison with the Gaussian models and experiments are
briefly discussed.
\end{abstract}


\section{Introduction} \label{sec:Intro}

In recent years, considerable progress has been achieved in the
understanding of intermittency and anomalous scaling of fluid
turbulence. Both the natural and numerical experiments suggest
that the deviation from the classical Kolmogorov theory
\cite{Monin,Legacy,Sree} is even more strongly pronounced for a
passively advected scalar field $\theta(x)\equiv\theta(t,{\bf x})$
(temperature, entropy, density of an impurity etc) than for the
velocity field itself; see e.g. Ref.~\cite{Sree,War,Ant,Helium}
and literature cited therein. At the same time, the problem of
passive advection appears more easily tractable theoretically:
even simplified models describing the advection by a ``synthetic''
velocity field ${\bf v}(x)\equiv\{v_i(x)\}$ with a given Gaussian
statistics reproduce many of the anomalous features of genuine
turbulent heat or mass transport observed in experiments.
Therefore, the problem of passive scalar advection, being of
practical importance in itself, may also be viewed as a starting
point in studying intermittency and anomalous scaling in the
turbulence on the whole.

The issue of interest is, in particular, the behavior of the equal-time
structure functions of the scalar field
\begin{equation}
S_{n}(r) =\big\langle[\theta(t,{\bf x})-\theta(t,{\bf x'})]^{n}
\big\rangle, \qquad {\bf r}={\bf x}-{\bf x'}, \quad r =|{\bf r}| .
\label{struc}
\end{equation}
The concept of anomalous scaling implies a power-law behavior of the
functions (\ref{struc}) in the inertial-convective range of scales,
$S_{n} \propto r^{\zeta_{n}}$, with a nonlinear dependence
of the exponents $\zeta_{n}$ on $n$; see
e.g. Refs.~\cite{Monin,Legacy,Sree,War,Ant,Helium}.

The crucial breakthrough in theoretical research
is related to a simple model of a passive scalar
quantity advected by a random Gaussian field, white in time and self-similar
in space, known as the Kraichnan's rapid-change model \cite{Kraich1}. There,
for the first time the existence of anomalous scaling was established on the
basis of a microscopic model \cite{Kraich2} and the corresponding anomalous
exponents were calculated within controlled approximations
\cite{GK,Falk1,Pumir,Siggia} and a systematic perturbation expansion in a
formal small parameter \cite{RG}. Detailed review of the recent theoretical
research on the passive scalar problem and the bibliography can be found in
Ref.~\cite{FGV}.

In the ``zero-mode approach,'' developed in \cite{GK,Falk1,Pumir,Siggia}
(see also \cite{FGV}), nontrivial anomalous exponents are related to the
zero modes (unforced solutions) of the closed exact differential equations
satisfied by the equal-time correlation functions. From the field theoretic
viewpoint, this is a realization of the well-known idea of self-consistent
(bootstrap) equations, which involve skeleton diagrams with dressed lines
and dropped bare terms. Owing to very special features of the rapid-change
models (linearity in the passive field and time decorrelation of the
advecting field) such equations are exactly given by one-loop approximations,
and the resulting equations in the coordinate space are differential (and not
integral or integro-differential as in the case of a general field theory).
In this sense, the model is ``exactly solvable.'' Furthermore, in contrast to
the case of nonzero correlation time, closed equations are obtained for the
equal-time correlations, which are Galilean invariant and, therefore,
not affected by the so-called ``sweeping effects'' that would obscure the
relevant physical interactions.

The second systematic analytical approach to the rapid-change
model, proposed in papers \cite{RG}, is based on the field
theoretic renormalization group (RG) and operator product
expansion (OPE). There, anomalous scaling emerges as a consequence
of the existence in the model of composite fields with negative
scaling dimensions (``dangerous composite operators''), identified
with the anomalous exponents. This allows one to give alternative
derivation of the anomalous scaling, to construct a systematic
perturbation expansion for the anomalous exponents, analogous to
the famous $\eps$ expansion in the RG theory of critical behavior
(see the monographs \cite{Zinn,Book3} and references therein), and
to calculate the exponents to the second \cite{RG,RG1} and third
\cite{kub,cube} orders.

The two approaches complement each other well: the zero-mode technique
allows for exact (nonperturbative) solutions for the anomalous exponents
related to second-order correlation functions \cite{Falk1,V96,LM} (they are
nontrivial for passive vector fields or  anisotropic sectors for scalar
fields), while the RG approach form the basis for systematic perturbative
calculations of the higher-order anomalous exponents. For anisotropic
velocity ensembles or/and passively advected vector fields, as well as for
passive advection of extended objects (polymers or membranes), where the
calculations become rather involved, all the existing results for
higher-order correlation functions were derived only by means of the RG
approach and only to the leading order in $\eps$
\cite{vektor,vektor1,vektor2,Wiese}.

From a more physical point of view, zero modes can be interpreted as
statistical conservation laws in the dynamics of particle clusters
\cite{slow}. The concept of statistical conservation laws appears rather
general, being also confirmed by numerical simulations of Refs.
\cite{CV,SCL}, where the passive advection in the two-dimensional
Navier--Stokes (NS) velocity field \cite{CV} and a shell model of a
passive scalar \cite{SCL} were studied. This observation is rather
intriguing because in those models no closed equations for
equal-time quantities can be derived due to the fact that the
advecting velocity has a finite correlation time (for a passive
field advected by a velocity with given statistics, closed
equations can be derived only for different-time correlation
functions, and they involve infinite diagrammatic series).

One may thus conclude that breaking the artificial assumptions of
the time decorrelation and Gaussianity of the velocity field is the
crucial point.

Besides the calculational efficiency, an important advantage of the RG
approach is its relative universality: it is not bound to the aforementioned
``solvability'' of the rapid-change model and can also be applied to the case
of finite correlation time or non-Gaussian advecting field. In Refs.
\cite{RG3,Juha2} (see also \cite{RG4} for the case of compressible
flow and \cite{RG5} for a passive vector field) the RG and OPE were applied
to the problem of a passive scalar advected by a Gaussian self-similar
velocity with finite (and not small) correlation time. The energy spectrum
of the velocity in the inertial range has the form
${\cal E}(k)\propto k^{1-2\eps}$, while the correlation time at the momentum
$k$ scales as $t(k)\propto k^{-2+\eta}$. It was shown that, depending on the
values of the exponents $\eps$ and $\eta$, the model reveals various types of
inertial-range scaling regimes with nontrivial anomalous exponents, which
were explicitly derived to the first \cite{RG3,RG4} and second \cite{Juha2}
orders of the double expansion in $\eps$ and $\eta$.
The most interesting case is $\eta=\eps$, when the exponents can
be nonuniversal through the dependence on the ratio of the velocity
correlation time and the turnover time of the passive scalar.

Earlier, a similar model was proposed and studied in detail (using
numerical simulations, in two dimensions) in \cite{OU}. Various
aspects of the transport and dispersion of particles in random
Gaussian self-similar velocity fields with finite correlation time
were also studied in Refs.
\cite{ShS,Falk3,Eyink,AM,AM2,shell,Fann,Horvai}.

As was pointed out in Ref. \cite{OU}, the Gaussian model with
finite correlation time suffers from the lack of Galilean
invariance and therefore misrepresents the self-advection of
turbulent eddies. It is well known that the different-time
correlations of the Eulerian velocity field are not self-similar,
as a result of these ``sweeping effects,'' and depend
substantially on the integral scale. It would be much more
appropriate to impose the scaling relations for ${\cal E}(k)$ and
$t(k)$ in the Lagrangian frame, but this is embarrassing due to
the daunting task of relating Eulerian and Lagrangian statistics
for a flow with a finite correlation time.\footnote{In this
connection, it should be noted that, due to the time
decorrelation, in the rapid-change model there is no problem in
relating Eulerian and Lagrangian statistics of the velocity field:
they are identical. This allows one to perform very accurate
numerical simulations in the Lagrangian frame; see
Refs.~\cite{VMF}.} In the RG and OPE formalism, the sweeping by
the large-scale eddies is related to the contributions of the
composite operators built solely of the velocity field ${\bf
v}(x)$ and its temporal derivatives, as discussed in detail in
Refs. \cite{Book3,JETP,UFN,turbo,Komp} for the case of the
stochastic NS equation. In the Gaussian model those operators
become dangerous (that is, their scaling dimensions become
negative) for $\eps\ge1/2$, which gives rise to strong infrared
divergences in the correlation functions \cite{RG3}. This means
that the sweeping effects, negligible for small $\eps$'s, become
important for $\eps\ge1/2$. In a Galilean-invariant model, such
operators give no contribution to the quantities like
(\ref{struc}), as explained in \cite{JETP,UFN,turbo,Komp} for the
NS case. In the Gaussian case, these IR divergences persist in the
structure functions, which provides not only an upper bound for
the reliability of the $\eps$--$\eta$ expansion, but also a
natural bound for the validity of the Gaussian model itself (which
excludes, in particular, the most realistic Kolmogorov's value
$\eps=4/3$ and its vicinity). These conclusions agree with the
nonperturbative analysis of Ref. \cite{Horvai}, where the value of
$\eps=1/2$ was reported as the threshold between two qualitatively
different regimes for a Lagrangian particle advected by a Gaussian
velocity ensemble. The same threshold value of $\eps=1/2$ was
obtained earlier in Refs. \cite{AM} for a two-dimensional strongly
anisotropic model.

In this paper, we shall study the case of a passive scalar field, advected
by a non-Gaussian velocity field governed by the stochastic NS
equation. To be precise, the advection-diffusion equation for the scalar
field has the form
\begin{equation}
\nabla_{t}\theta =\kappa _0\partial^{2} \theta+f, \qquad
\nabla_{t} \equiv \partial _t + (v_{i}\partial_{i}),
\label{1}
\end{equation}
where $\partial _t \equiv \partial /\partial t$, $\partial_i
\equiv \partial /\partial x_{i}$, $\nabla_{t}$ is the Lagrangian
derivative, $\kappa_0$ is the thermal conductivity or molecular
diffusivity, $\partial^{2}$ is the Laplace operator, and $f\equiv
f(x)$ is an artificial Gaussian random noise with zero mean and
correlation function
\begin{equation}
\langle  f(x)  f(x')\rangle =  \delta(t-t') C({\bf r}), \qquad
{\bf r}={\bf x}-{\bf x'}.
\label{2}
\end{equation}
The detailed form of the function $C$ is unessential; it is only
important that $C$ decreases rapidly for $r\gg L$, where $L$ is
some integral scale. The noise maintains the steady state of the
system and, if $C$ depends on the vector ${\bf r}$ and not only on
its modulus $r=|{\bf r}|$, is a source of large-scale anisotropy.
With no loss of generality it can be assumed that the function $C$
is dimensionless. The absence of the time correlation in (\ref{2})
is also unessential; in a more realistic formulation, the noise
can be replaced by an imposed constant gradient of the scalar
field; see Refs. \cite{Pumir,Siggia,RG3,RG4,OU}.

The transverse (divergence-free, due to the incompressibility
condition $\partial_i v_i=0$) velocity field satisfies the NS
equation with a random driving force
\begin{equation}
\nabla _t v_i=\nu _0\partial^{2} v _i-\partial _i {\cal P}+f_i,
\label{1.1}
\end{equation}
where ${\cal P}$ and $f_i$ are the pressure and the transverse random force
per unit mass (all these quantities depend on $x$). We assume for $f$ a
Gaussian distribution with zero mean and correlation function
\begin{equation}
\big\langle f_i(x)f_j(x')\big\rangle = \frac{\delta (t-t')}{(2\pi)^{d}}\,
\int_{k\ge m} d{\bf k}\, P_{ij}({\bf k})\, d_f(k)\, \exp \big[{\rm i}{\bf k}
\left({\bf x}-{\bf x}'\right)\big] ,
\label{1.2}
\end{equation}
where $P_{ij}({\bf k}) =\delta _{ij}  - k_i k_j / k^2$ is the transverse
projector, $d_f(k)$ is some function of $k\equiv |{\bf k}|$ and model
parameters, and $d$ is the dimension of the ${\bf x}$ space.
The momentum $m=1/\ell$, the reciprocal of another integral scale $\ell$,
provides IR regularization. Its precise form is unessential, the sharp
cutoff is the most convenient choice from purely calculational reasons.
For definiteness, in what follows it is always assumed that $L\gg \ell$,
that is, the largest scale in the problem is the integral scale related
to the scalar noise; it will be set to infinity whenever possible.

The standard RG formalism is applicable to the problem (\ref{1.1}),
(\ref{1.2}) if the correlation function of the random force is chosen
in the power form
\begin{equation}
d_f(k)=D_0\,k^{4-d-2\varepsilon},
\label{1.9}
\end{equation}
where $D_{0}>0$ is the positive amplitude factor and the exponent
$0<\eps\le 2$ plays the role analogous to that played by $4-d$ in
the RG theory of critical behavior \cite{Zinn,Book3}. The form
(\ref{1.9}) is widely used in the RG theory of turbulence since
the pioneering works \cite{Nelson,Dominicis,Frisch,Pismak}. The
most realistic value of the exponent is $\eps=2$: with an
appropriate choice of the amplitude, the function (\ref{1.9}) for
$\eps\to2$ turns to the delta function, $d_f(k) \propto
\delta({\bf k})$, which corresponds to the injection of energy to
the system owing to interaction with the largest turbulent eddies;
for a more detailed discussion see Refs.
\cite{Book3,UFN,turbo,Komp}.

The results of the RG analysis of the model
(\ref{1.1})--(\ref{1.9}) are reliable and internally consistent
for small $\eps$, while the possibility of their extrapolation to
the real value $\eps=2$ and thus their relevance for the real
fluid turbulence is far from obvious; see e.g. Ref. \cite{Komp}
for a recent discussion. We shall not discuss this important
problem in detail and restrict ourselves with a few remarks which
will be relevant in what follows.

The time decorrelation of the random force guarantees that the
full stochastic problem (\ref{1})--(\ref{1.2}) is Galilean
invariant for all values of the model parameters, including $D_0$
and $\eps$. As a consequence, the ordinary perturbation theory for
the model (that is, the expansion in the nonlinearities, or,
equivalently, in $D_0$ from (\ref{1.9})), is manifestly Galilean
covariant: all the exact relations between the correlation
functions imposed by the Galilean symmetry (Ward identities) are
satisfied order by order. The renormalization procedure does not
violate the Galilean symmetry, so that the improved perturbation
expansion, obtained with the aid of RG and OPE, remains covariant.
This means, in particular, that the Galilean invariant quantities,
for example, the equal-time structure functions (\ref{struc}), are
not affected by the sweeping (here, the latter becomes important
for $\eps\ge3/2$ \cite{JETP}). More formally, the contributions of
the ``dangerous'' operators built of the velocity field and its
temporal derivatives, do not appear in the OPE for invariant
correlation functions; see Refs. \cite{Book3,JETP,UFN,turbo,Komp}
for detailed discussion.

This means that the scaling relations, obtained for Galilean
invariant quantities for small $\eps$, in model
(\ref{1})--(\ref{1.9}) can be extrapolated beyond the threshold
$\eps=3/2$ despite the fact that the sweeping becomes important
there. More physically, this means that, in contrast to the
Gaussian model, the relative motion of the fluid or impurity
particles in the inertial-convective range of scales is not
affected by overall sweeping by the large-scale eddies. Indeed,
the most recent numerical simulations of the model
(\ref{1.1})--(\ref{1.9}) have shown that the scaling relations,
obtained by the RG analysis for the structure functions, remain
valid for $\eps$ as high as $\eps=7/4$ \cite{Biferale} (see also
an earlier work \cite{Pandit}).

For small $\eps$, critical dimensions of all composite operators in the
model (\ref{1.1})--(\ref{1.9}) are positive. As a result, the scaling
behavior of the velocity correlation functions is not anomalous, in the
sense that they have a finite limit for $\ell=1/m\to\infty$, and the
corresponding scaling exponents are multiples of a single quantity
(critical dimension of the velocity field).

However, the same numerical simulations \cite{Biferale,Pandit}
suggest that, as $\eps$ increases, the behavior of the model
(\ref{1.1})--(\ref{1.9}) undergoes a qualitative changeover and
the scaling becomes anomalous, in the sense that the exponents of
the structure functions become different from the results of naive
extrapolation of the small-$\eps$ prediction and, probably, become
independent of $\eps$. We shall return to this important issue in
the Conclusion, and in the bulk of the paper we shall concentrate
on the behavior of the passive scalar field in the model
(\ref{1})--(\ref{1.9}), which appears highly nontrivial.

We will show that, already for infinitesimal values of $\eps$, when the
velocity statistics is not yet intermittent, the scalar field, advected
by such velocity ensemble, displays anomalous scaling behavior.
The corresponding anomalous exponents can be calculated within a
systematic perturbation expansion, as series in $\eps$.

The plan of the paper is the following.

Detailed description of the model has already been given above. In
Sec.~\ref{sec:QFT} we give the field theoretic formulation of the
original stochastic problem and present the corresponding
diagrammatic technique. In Sec.~\ref{sec:RGE} we analyze UV
divergences in the model, establish its multiplicative
renormalizability, and derive the corresponding RG equations. In
Sec.~\ref{sec:FPD} we show that the RG equations of our model have
the only IR attractive fixed point in the physical range of
parameters; its coordinates are calculated to the second order of
the $\eps$ expansion. Existence of such fixed point means that the
correlation functions of our model in the IR range ($1/r \sim k
\sim m \ll \Lambda$) exhibit scaling behavior with certain
critical dimensions $\Delta_{F}$ of all the fields and parameters
$F$. This determines the dependence of the correlation functions
on the argument $\Lambda r$ (but not yet on $mr$). In general, the
dimensions are calculated as series in $\eps$, but for some basic
quantities (velocity and scalar fields, their powers, and
frequency) they are found exactly.

The key role in the following is played by the composite operators
built of the gradients of the scalar field. They are introduced in
Sec.~\ref{sec:Operators}, and the corresponding dimensions
$\Delta_{F}$ are given to the first order in $\eps$ (one-loop
approximation). In Sec.~\ref{sec:OPE} we introduce the
operator-product expansion and demonstrate its relevance to the
issue of inertial-range anomalous scaling. We show that the
critical dimensions of aforementioned composite operators can be
identified with the anomalous exponents, which describe the
dependence of the correlation functions on the argument $mr$. The
scalar operators are ``dangerous'' ($\Delta_{F}<0$), which implies
anomalous scaling (singular dependence on $mr$ and divergence for
$mr\to0$). The anomalous exponents of anisotropic contributions
are determined by the critical dimensions of tensor composite
fields and thus they also exhibit nontrivial scaling behavior.

The largest Sec.~\ref{sec:Dimensions} is devoted to the
calculation of the anomalous exponents (critical dimensions of the
composite operators built of the gradients of the scalar field) to
the order $\eps^{2}$ (two-loop approximation). The results look
rather cumbersome and are presented in a separate section,
Sec.~\ref{sec:Exponents}. The discussion of the results, their
relevance for the real passive advection, comparison with the
Gaussian models and experiments are given in
Sec.~\ref{sec:Conclusion}.

\section{Field theoretic formulation} \label{sec:QFT}

The field theoretic formulation and renormalization of the problem
(\ref{1})--(\ref{1.9}) is discussed in detail in Ref.
\cite{Book3,UFN,turbo}; below we confine ourselves to only the
necessary information.

According to the general theorem \cite{MSR}, stochastic problem
(\ref{1})--(\ref{1.9}) is equivalent to the field
theoretic model of the doubled set of fields
$\Phi\equiv\{{\bf v}', {\bf v}, \theta', \theta\}$ with action functional
\begin{equation}
S(\Phi )= S_{v}({\bf v}', {\bf v}) + \theta' D_{\theta} \theta'/2+
\theta' \left[ -\nabla_{t}+\kappa_{0}\partial^{2}
\right] \theta,
\label{action}
\end{equation}
where
\begin{equation}
S_{v}({\bf v}', {\bf v}) =
v 'D_{v}v'/2+ v'\left[-\nabla_t +\nu _0\partial^{2} \right]v
\label{actionV}
\end{equation}
is the action functional for the stochastic problem (\ref{1.1})--(\ref{1.9}),
$D_{\theta}$ and $D_{v}$ are the correlation functions (\ref{2}) and
(\ref{1.2}) of the random forces $f$ and $f_{i}$, respectively, and all the
required integrations over $x=\{t,{\bf x}\}$ and summations over the vector
indices are understood, for example,
$$ v'(v\partial)v \equiv \int dt \int d{\bf x} \ v_{i}'
(v_{j}\partial_{j}) v_{i}'. $$ The auxiliary vector field is also
transverse, $\partial_{i}v_{i}'=0$, which allows to omit the
pressure term on the right-hand side of Eq. (\ref{actionV}), as
becomes evident after the integration by parts:
$$ \int dt \int d{\bf x} \ v_{i}'\partial_{i} {\cal P} = - \int dt
\int d{\bf x} \ {\cal P} (\partial_{i}v_{i}') =0 . $$ Of course,
this does not mean that the pressure contribution can simply be
neglected: the field $v'$ acts as the transverse projector and
selects the transverse part of the expressions to which it is
contracted in Eq.~(\ref{actionV}).

Formulation (\ref{action}), (\ref{actionV}) means that statistical
averages of random quantities in the original stochastic problem
can be represented as functional averages with the weight $\exp
S(\Phi)$, and the generating functionals of total [$G(A)$] and
connected [$W(A)$] correlation functions of the problem are
represented by the functional integral
\begin{equation}
G(A)=\exp  W(A)=\int {\cal D}\Phi \exp [S(\Phi )+A\Phi ]
\label{14}
\end{equation}
with arbitrary sources $A\equiv \{
A^{v'},A^{v},A^{\theta'},A^{\theta}\}$ in the linear form $A\Phi
\equiv \sum _{\Phi}\int dx \,A^{\Phi}(x)\Phi (x)$.

The model (\ref{actionV}) corresponds to a standard Feynman
diagrammatic technique; the bare propagators (lines in the diagrams)
in the frequency--momentum ($\omega$--$\k$) representation have the forms
\begin{eqnarray}
\bigl\langle v_{i} v_{j}' \bigr\rangle _0 =
\bigl\langle v_{i}' v_{j} \bigr\rangle _0^{*}= (-{\rm i}\omega+\nu_0
k^{2})^{-1}\, P_{ij}({\bf k}), \qquad
\bigl\langle v_{i} v_{j} \bigr\rangle _0 = (\omega^{2}+\nu_0^{2}k^{4})^{-1}\,
d_{f}(k) P_{ij}({\bf k}), \qquad  \bigl\langle v_{i}' v_{j}' \bigr\rangle _0
= 0
\label{linesV}
\end{eqnarray}
with $d_{f}(k)$ from Eq. (\ref{1.9}). The interaction in (\ref{actionV})
corresponds to the triple vertex $-v'(v\partial)v= v'_{i}V_{ijs}v_{j}v_{s}/2$
with vertex factor
\begin{equation}
V_{ijs} = {\rm i} (k_{j}\delta_{is}+k_{s}\delta_{ij}),
\label{vertexV}
\end{equation}
where ${\bf k}$ is the momentum argument of the field $v'$. The full
problem (\ref{action}) involves additional propagators
\begin{eqnarray}
\bigl\langle \theta\theta' \bigr\rangle _0 =
(-{\rm i}\omega + \kappa_{0}k^{2})^{-1}, \qquad
\bigl\langle \theta\theta \bigr\rangle _0 = C(\k)\,
(\omega^{2}+ \kappa_{0}^{2}k^{4})^{-1},  \qquad
\bigl\langle \theta'\theta' \bigr\rangle _0 =0,
\label{lines}
\end{eqnarray}
where $C(\k)$ is the Fourier transform of the function $C({\bf r})$
from (\ref{2}); the additional vertex factor in
$\theta'(v\partial) \theta = \theta' V_{i} v_{i}\theta$ has the form
\begin{equation}
V_{i} = {\rm i} k_{i} ,
\label{vertex}
\end{equation}
where ${\bf k}$ is the momentum of the field $\theta'$.

\section{Renormalization and RG equations} \label{sec:RGE}

The analysis of UV divergences is based on the analysis of
canonical dimensions. In contrast to static models, dynamical
models of the type (\ref{action}), (\ref{actionV}) have two
scales, i.e., the canonical dimension of some quantity $F$ (a
field or a parameter in the action functional) is described by two
numbers, the momentum dimension $d_{F}^{k}$ and the frequency
dimension $d_{F}^{\omega}$. They are determined so that $[F] \sim
[L]^{-d_{F}^{k}} [T]^{-d_{F}^{\omega}}$, where $L$ is the length
scale and $T$ is the time scale. The dimensions are found from the
obvious normalization conditions $d_k^k=-d_{\bf x}^k=1$,
$d_k^{\omega } =d_{\bf x}^{\omega }=0$, $d_{\omega }^k=d_t^k=0$,
$d_{\omega }^{\omega }=-d_t^{\omega }=1$, and from the requirement
that each term of the action functional be dimensionless (with
respect to the momentum and frequency dimensions separately).
Then, based on $d_{F}^{k}$ and $d_{F}^{\omega}$, one can introduce
the total canonical dimension $d_{F}=d_{F}^{k}+2d_{F}^{\omega}$
(in the free theory, $\partial_{t}\propto\partial^{2}$), which
plays in the theory of renormalization of dynamical models the
same role as the conventional (momentum) dimension does in static
problems.

The canonical dimensions for the problem (\ref{action}),
(\ref{actionV}) are summarized in Table \ref{table1}, where we
introduced the new parameters (``coupling constants'' or
``charges'')
\begin{equation}
g_{0}\equiv D_{0}/\nu_0^3, \quad u_{0}\equiv \kappa_{0}/\nu_0
\label{Prad}
\end{equation}
instead of $D_{0}$ and $\kappa_{0}$ and included the dimensions of
renormalized parameters, which will appear later on. The dimensionless
ratio $u_{0}$ has the meaning of the reciprocal of the Prandtl number. From
Table~\ref{table1} it follows that the model (\ref{action}), (\ref{actionV})
is logarithmic (the coupling constant $g_0$ is dimensionless) at $\eps=0$,
and the UV divergences have the form of the poles in $\eps$ in the
correlation functions of the fields $\Phi$.

The total canonical dimension of an arbitrary 1-irreducible correlation
function $\Gamma=\langle\Phi\cdots\Phi\rangle_{\rm 1-ir}$ is given by
the relation $d_{\Gamma }=d_{\Gamma }^k+2d_{\Gamma }^{\omega }= (d+2)-
\sum_{\Phi}N_{\Phi }d_{\Phi}$, where $N_{\Phi}=\{N_{v}',N_{v}, N_{\theta'},
N_{\theta}\}$ are the numbers of corresponding fields entering into the
function $\Gamma$, and the summation over all types of the fields is implied.
The total dimension $d_{\Gamma}$ plays the role of the formal index of the UV
divergence: superficial UV divergences, whose removal requires counterterms,
can be present only in those functions $\Gamma$ for which $d_{\Gamma}$ is a
non-negative integer. Analysis of the divergences in our model should be
augmented by the following considerations:

(i) For any model with the Martin--Siggia--Rose-type action, that is, the
action of the form (\ref{action}), (\ref{actionV}), all the 1-irreducible
functions with $N_{{\bf v}'}=N_{\theta'}=0$ contain closed circuits of
retarded propagators and vanish.

(ii) If for some reason a number of external momenta occurs as an overall
factor in all the diagrams of a given Green function, the real index of
divergence $d_{\Gamma}'$ is smaller than $d_{\Gamma}$ by the corresponding
number of unities. The correlation function requires counterterms only if
$d_{\Gamma}'$  is a non-negative integer. In our model, the derivative
$\partial$ at the vertices $v'(v\partial)v$ and $\theta'(v\partial) \theta$
can be moved onto the fields $v'$ and $\theta'$ using the integration by
parts, by virtue of the transversality of the field $v$. This decreases the
real index of divergence: $d_{\Gamma}' = d_{\Gamma}- N_{v'}- N_{\theta'}-
N_{\theta}$, and the fields $v'$, $\theta'$ and $\theta$ enter the
counterterms only in the form of the derivatives, $\partial v'$ and so on.

(iii) From the explicit form of the vertex and bare propagators it follows
that $N_{\theta'}- N_{\theta}=2N_{0}$ for any 1-irreducible function, where
$N_{0}\ge0$ is the total number of bare propagators $\langle \theta \theta
\rangle _0$ entering into the function (obviously, no function with $N_{0}<0$
can be constructed). Therefore, the difference $N_{\theta'}- N_{\theta}$ is
an even non-negative integer for any nonvanishing function. This is a
consequence of the linearity of the original stochastic equation (\ref{1})
in the field $\theta$.

(iv) Galilean symmetry of our problem requires that the counterterms to
the action be invariant. In particular, the monomials $v'\partial_{t}v$,
$v'(v\partial)v$, $\theta'\partial_{t}\theta$, and $\theta'(v\partial)\theta$
can appear only in the form of covariant derivatives $v'\nabla_{t}v$ and
$\theta'\nabla_{t}\theta$.

From Table \ref{table1} we find
$d_{\Gamma}=(d+2) -(d-1)N_{v'} -N_{v}+N_{\theta} -(d+1)N_{\theta'}$ and
$d_{\Gamma}'=(d+2) -d N_{v'} -N_{v}-(d+2)N_{\theta'}$.
Bearing in mind that $N_{\theta'} \ge N_{\theta}$ we find that superficial
divergences can only be present in the 1-irreducible functions
$\langle v'v\rangle_{\rm 1-ir}$ and $\langle\theta'\theta\rangle_{\rm 1-ir}$,
and the corresponding counterterms reduce to the forms $v'\partial^{2}v$ and
$\theta'\partial^{2}\theta$. The monomials $v'\partial_{t}v$ and
$\theta'\partial_{t}\theta$ do not contain spatial derivatives and therefore
they are ruled out by the property (ii). Then the property (iv) excludes
the monomials $v'(v\partial)v$ and $\theta'(v\partial)\theta$ (allowed by
dimensional considerations).

In the special case $d=2$ a new UV divergence appears in the
1-irreducible function $\langle v'v'\rangle_{\rm 1-ir}$. This case
requires special attention, see Refs. \cite{Two}, and from now on
we always assume $d>2$. Then the inclusion of the counterterms is
reproduced by the multiplicative renormalization of the action
functional (\ref{action}), (\ref{actionV}) with only two
independent renormalization constants $Z_{1,2}$:
\begin{equation}
S_{R}(\Phi)= S_{vR}({\bf v}', {\bf v}) + \theta' D_{\theta} \theta'/2+
\theta' \left[ -\nabla_{t}+ u\nu Z_{2} \partial^{2}
\right] \theta,
\label{Rac}
\end{equation}
and
\begin{equation}
S_{vR}({\bf v}', {\bf v}) =
v 'D_{v} v'/2+ v'\left[-\nabla_t +\nu Z_{1}\partial^{2} \right]v.
\label{RaV}
\end{equation}

In the one-loop approximation the renormalization constants have the forms
\begin{equation}
Z_{1}= 1 -  \frac{g \bar S_{d}\, (d-1)}{8(d+2)\eps}+O(g^{2}), \qquad
Z_{2}= 1 -  \frac{g \bar S_{d}\, (d-1)}{4du(u+1)\eps }+O(g^{2}),
\label{ZZ}
\end{equation}
where $\bar S_{d}=S_{d}/(2\pi)^{d}$ and $S_d = 2\pi ^{d/2}/\Gamma
(d/2)$ is the surface area of the unit sphere in $d$-dimensional
space. Here and below, we use the minimal subtraction (MS) scheme,
in which all renormalization constants have the forms ``1 + only
poles in $\eps$.'' Since the velocity field is not affected by the
fields $\theta$ and $\theta'$, the constant $Z_{1}$ is independent
of $u$; the one-loop expression (\ref{ZZ}) was presented in
\cite{Pismak} and the two-loop calculation was performed much
later in Refs. \cite{Komp,APS}. The constant $Z_{2}$ is determined
by the 1-irreducible function $\langle\theta'\theta\rangle_{\rm
1-ir}$, which does not involve the correlation function (\ref{2});
see item (iii) above. Therefore, $Z_{2}$ in our model coincides
exactly with the corresponding renormalization constant for the
case of a passive scalar without the random noise, derived in the
one-loop approximation in Ref. \cite{AVH} (see also Refs.
\cite{Yakhot2}).\footnote{In the books \cite{Book3,turbo}, there
is a misprint in the expression for $Z_{2}$ on pages 709 and 115,
respectively. It is also interesting to note that the one-loop
expression for this constant coincides with its analog for a
passive magnetic field (``kinetic regime'') derived in Refs.
\cite{MHD}.}

Renormalization (\ref{Rac}), (\ref{RaV}) can be reproduced by the
following multiplicative renormalization of the parameters
$g_{0}$, $u_{0}$, and $\nu_{0}$:
\begin{equation}
g_{0}= g\mu^{2\eps} Z_{g}, \quad \nu_0=\nu Z_{\nu}, \quad
u_0=u Z_{u}, \qquad Z_{\nu}=Z_{1}, \quad Z_{u}=Z_{2}Z_{1}^{-1}, \quad
Z_{g}=Z_{1}^{-3}.
\label{renorm}
\end{equation}
Here $g$, $u$, and $\nu$ (without a subscript) are the
renormalized analogs of the corresponding bare parameters (with
the subscript 0) and $\mu$ is the reference mass (additional
arbitrary parameter of the renormalized theory). The last relation
in (\ref{renorm}) is the consequence of the absence of
renormalization of the term with $D_{v}$ in (\ref{actionV}). The
amplitude $D_{0}$ in the term with $D_{v}$ should be expressed in
renormalized parameters using the relations $D_{0}=g_{0}\nu_0^{3}=
g\mu^{2\eps}\nu^3$. No renormalization of the fields $\Phi$ and
``masses'' $m=1/\ell$, $1/L$ is needed.

Let $W(e_{0})$ be some correlation function in the original model
(\ref{action}) and $W_{R}(e,\mu)$ its analog in the renormalized
theory with action (\ref{Rac}). Here $e_{0}$ is the complete set
of bare parameters, and $e$ is the set of their renormalized
counterparts. The relation $S(\Phi, e_{0}) =S_{R}(\Phi,e,\mu)$ for
the action functionals yields $W(e_{0})=W_{R}(e,\mu)$ for any
correlation function of the fields $\Phi$; the only difference is
in the choice of variables and in the form of perturbation theory
(in $g$ instead of $g_{0}$). We use $\widetilde{\cal D}_{\mu}$ to
denote the differential operation $\mu\partial_{\mu}$ for fixed
$e_{0}$ and operate on both sides of this equation with it. This
gives the basic RG equation:
\begin{equation}
{\cal D}_{RG}\,W_{R}(e,\mu) = 0,  \qquad
{\cal D}_{RG}\equiv {\cal D}_{\mu} + \beta_{g}\partial_{g} +
\beta_{u}\partial_{u} -\gamma_{\nu}{\cal D}_{\nu}.
\label{RG1}
\end{equation}
Here ${\cal D}_{RG}$ is the operation $\widetilde{\cal D}_{\mu}$ expressed
in the renormalized variables, ${\cal D}_{x}\equiv x\partial_{x}$ for any
variable $x$, and the RG functions (the $\beta$ functions and the anomalous
dimensions $\gamma$) are defined as
\begin{eqnarray}
\gamma_{F} \equiv \widetilde{\cal D}_{\mu} \ln Z_{F}
\quad {\rm for\ all\ } F,
\label{RGF}
\end{eqnarray}
\begin{eqnarray}
\beta_{g} \equiv \widetilde{\cal D}_{\mu} g = g\left[-2\eps+3\gamma_{\nu}
\right], \qquad \beta_{u} \equiv \widetilde{\cal D}_{\mu} u =
-u \gamma_{u},
\label{RGF2}
\end{eqnarray}
where the relations (\ref{renorm}) have been used.

\section{Fixed point, infrared scaling, and critical dimensions}
\label{sec:FPD}

It is well known that possible scaling regimes of a renormalizable model are
associated with the IR stable fixed points of the corresponding RG equations.
In our model, the coordinates $g_{*},u_{*}$ of the fixed points are found
from the equations
\begin{equation}
\beta_{g} (g_{*})=\beta_{u} (g_{*},u_{*})=0
\label{points}
\end{equation}
with the $\beta$ functions from (\ref{RGF2}), while the type of a point is
determined by the $2\times2$ matrix consisting of the elements $\Omega=\{
\partial_{g}\beta_{g}, \partial_{u}\beta_{g}, \partial_{g}\beta_{u},
\partial_{u}\beta_{u}\}$ calculated at the point $g_{*},u_{*}$.
For IR stable fixed points the matrix $\Omega$ is positive, i.e., the real
parts of all its eigenvalues are positive. In our model $\partial_{u}\beta_{g}$
vanishes identically, and the eigenvalues of the matrix $\Omega$ are simply
given by its diagonal elements.

The analysis of the explicit one-loop expressions shows that, for $\eps>0$,
the RG equations for our model possesses the only IR attractive fixed point
in the physical range of parameters ($g_{*},u_{*}$ must be positive). The
coordinates are calculated as series in $\eps$,
\begin{equation}
g_{*}= g_{*}^{(1)}\eps+g_{*}^{(2)}\eps^{2}+O(\eps^{3}), \qquad
u_{*}= u_{*}^{(0)}+u_{*}^{(1)}\eps+O(\eps^{2}),
\label{FPE}
\end{equation}
with the one-loop approximation
\begin{equation}
g_{*}^{(1)}\bar S_{d}= \frac{8(d+2)}{3(d-1)},
\qquad
u_{*}^{(0)}=\frac{1}{2}\left(\sqrt{1+8(d+2)/d}-1\right)
\label{FP1}
\end{equation}
(with $\bar S_{d}$ from (\ref{ZZ})) for arbitrary $d>2$. The
two-loop result for  $g_{*}$ was presented in \cite{Komp,APS}:
$g_{*}^{(2)}\approx -1.01 g_{*}^{(1)}$ for $d=3$. We have also
calculated the two-loop correction for $u_{*}$:
$u_{*}^{(1)}\approx -0.035 u_{*}^{(0)}$ for $d=3$. We shall not
expound on the derivation of this result because, as we shall see
below, the two-loop corrections to the coordinates $g_{*},u_{*}$
are not needed for the two-loop calculation of the anomalous
exponents.

Existence of the IR stable fixed point implies that the
correlation functions of our model in the IR range exhibit scaling
behavior with definite critical dimensions $\Delta_{F}$ of all the
fields and parameters $F$. Let $F$ be some multiplicatively
renormalized quantity (a field, parameter or composite operator),
that is, $F=Z_{F}F_{R}$ with certain renormalization constant
$Z_{F}$. Then its critical dimension is given by the expression
(see e.g. \cite{Book3,UFN,turbo})
\begin{equation}
\Delta[F]\equiv\Delta_{F} = d_{F}^{k}+ \Delta_{\omega}
d_{F}^{\omega}+\gamma_{F}^{*},  \qquad
\Delta_{\omega}=2-\gamma^{*}_{\nu},
\label{32B}
\end{equation}
where $d_{F}^{k}$ and $d_{F}^{\omega}$ are the corresponding
canonical dimensions, $\gamma_{F}^{*}$ is the value of the
anomalous dimension $\gamma_{F}(g)\equiv \widetilde{\cal D}_\mu
\ln Z_{F}$  at the fixed point in question, and $\Delta_{\omega}$
is the critical dimension of frequency. Owing to the exact
relation between $\gamma_{\nu}$ and $\beta_{g}$ in (\ref{RGF2}),
its value at the critical point is found exactly:
$\Delta_{\omega}=2-2\eps/3$ (without corrections of order
$\eps^{2}$ and higher). As a consequence, the critical dimensions
of some basis parameters, fields and composite operators are also
found exactly:
\begin{eqnarray}
\Delta[v^{n}]=n\Delta_{v}=n(1-2\eps/3), \qquad
\Delta[\theta^{n}]=n\Delta_{\theta}=n(-1+\eps/3),
\nonumber \\
\Delta[v']=d-\Delta_{v}, \qquad
\Delta[\theta']= d-\Delta_{\theta}, \qquad \Delta_{m}=1.
\label{Deltas}
\end{eqnarray}
To avoid possible misunderstanding, it should be noted that simple
linear relations for the dimensions of composite fields $v^{n}$
and $\theta^{n}$ follow from the fact that these operators are not
renormalized ($Z_{F}=1$). For the powers of the velocity, this is
a consequence of the Galilean symmetry (see
\cite{JETP,UFN,turbo}), while for $\theta^{n}$ will be discussed
below.

Let $G(r)=\langle F_{1}(x)F_{2}(x')\rangle$ be, for definiteness,
some equal-time two-point quantity, for example, the pair
correlation function of the primary fields $\Phi$ or some
multiplicatively renormalizable composite operators. The existence
of a nontrivial IR stable fixed point implies that in the IR
asymptotic region $\Lambda r \gg 1$ and any fixed $mr$ the
function $G(r)$ takes on the form
\begin{equation}
G(r) \simeq  \nu_{0}^{d_{G}^{\omega}}\, \Lambda^{d_{G}}
(\Lambda r)^{-\Delta_{G}}\, \xi(mr).
\label{RGR}
\end{equation}
Here the UV momentum scale $\Lambda$ is defined by the relations
$g_{0} = D_{0}/\nu_0^3 = \Lambda^{2\eps}$, and $\xi$ is certain
scaling function whose explicit form is not determined by the RG
equation itself. The canonical dimensions $d_{G}^{\omega}$,
$d_{G}$ and the critical dimension $\Delta_{G}$ ot the function
$G(r)$ are equal to the sums of the corresponding dimensions of
the quantities $F_{1,2}$.

\section{Renormalization of relevant composite operators}
\label{sec:Operators}

In the following, an important role will be played by the composite
operators built of the field $\theta(x)$ and its spatial derivatives.

We recall that the term ``local composite operator''  refers to any monomial
or polynomial built of the fields $\Phi$ and their derivatives at a single
spacetime point $x=\{t,{\bf x}\}$, for example $\theta^{n}$ or
$\theta'(v\partial)\theta$.

Coincidence of the field arguments in correlation functions containing an
operator $F$ gives rise to additional UV divergences, removed by a special
renormalization procedure. Owing to the renormalization, the critical
dimension $\Delta_{F}$ associated with certain operator $F$ is not in
general equal to the
simple sum of critical dimensions of the fields and derivatives entering
into $F$. As a rule, composite operators ``mix'' in renormalization,
that is, an UV finite renormalized operator is a linear combination
of unrenormalized operators, and vice versa.

In general, counterterms to a given operator $F$ are determined by all
possible 1-irreducible Green functions with one operator $F$ and arbitrary
number of primary fields,
$\Gamma=\langle F(x) \Phi(x_{1})\dots\Phi(x_{2})\rangle_{\rm 1-ir}$.
The total canonical dimension (formal index of divergence)
for such quantities is given by
\begin{equation}
d_{\Gamma} = d_{F} - \sum_{\Phi}N_{\Phi}d_{\Phi},
\label{index}
\end{equation}
with the summation over all types of fields entering into the function
and the canonical dimensions from Table \ref{table1}. For superficially
divergent diagrams, $d_{\Gamma}$ is a non-negative integer.

Consider the simplest operators of the form $\theta^{n}(x)$ with the
canonical dimension $d_{F}=-n$, entering into the structure functions
(\ref{struc}). From Table \ref{table1} and Eq. (\ref{index}) we obtain
$d_{\Gamma} = -n+N_{\theta}-(d-1)N_{v'}-N_{v} -(d+1)N_{\theta'}$, and
from the analysis of the diagrams it follows that the total number of
the fields $\theta$ entering into the function $\Gamma$ can never exceed
the number of the fields $\theta$ in the operator $\theta^{n}$ itself:
$N_{\theta}\le n$ (a consequence of the linearity of the original stochastic
equations in $\theta$). Therefore, the divergence can only exist in the
functions with $N_{v}=N_{v'}=N_{\theta'}=0$, and arbitrary value of
$n=N_{\theta}$, for which the formal index vanishes, $d_\Gamma =0$. However,
at least one of $N_{\theta}$ external ``tails'' of the field $\theta$ is
attached to a vertex $\theta'(v\partial)\theta$ (it is impossible to
construct nontrivial, superficially divergent diagram of the desired type
with all the external tails attached to the vertex $F$), at least one
derivative $\partial$ appears as an extra factor in the diagram, and,
consequently, the real index of divergence is necessarily negative.

This means that the operator $\theta^{n}$ requires no counterterms at all,
that is, it is in fact UV finite: $\theta^{n}=Z\,[\theta^{n}]^{R}$ with
$Z=1$. It then follows that the critical dimension of $\theta^{n}(x)$ is
simply given by the expression (\ref{32B}) with no correction from
$\gamma_{F}^{*}$ and therefore reduces to the sum of the critical dimensions
of the factors: $\Delta [\theta^{n}] = n\Delta_{\theta} =n (-1+\eps/3)$, as
already stated in Eq. (\ref{Deltas}).

Now let us turn to the scalar operators
\begin{equation}
F_{n} = (\partial_{i}\theta \partial_{i}\theta)^{n}
\label{Fn}
\end{equation}
with $d_{F}=0$, $d_{F}^{\omega}=-n$. As we shall see below, it is
their critical dimensions that determine the anomalous exponents
for the structure functions (\ref{struc}) and other quantities. In
this case, from Table \ref{table1} and Eq. (\ref{index}) we find
$d_\Gamma = N_{\theta}-N_{\bf v}-(d-1)N_{v'} -(d+1)N_{\theta'}$,
with the necessary condition $N_{\theta}\le 2n$ following from the
structure of the diagrams. It is also obvious from the analysis of
the diagrams that the counterterms to these operators can involve
the fields $\theta$, $\theta'$ only in the form of derivatives,
$\partial\theta$, $\partial\theta'$, so that the real index of
divergence $d_\Gamma'$ has the form $d_\Gamma' = d_\Gamma
-N_{\theta}-N_{\theta'}-N_{v'}= -N_{v} -(d+2)N_{\theta'}-dN_{v'}$.
It then follows that superficial divergences exist only in the
correlation functions with $N_{v}=N_{v'}=N_{\theta'}=0$ and any
$N_{\theta}\le 2n$, and the corresponding operator counterterms
are reduced to the form $F_{k}$ with $k\le n$. Therefore, the
operators $F_{n}$ can mix only with each other in renormalization,
the corresponding infinite renormalization matrix
$Z_{F}=\{Z_{nk}\}$ is in fact triangular, $Z_{nk}=0$ for $k>n$,
and the critical dimensions associated with the operators $F_{n}$
are determined by the diagonal elements $Z_{n}\equiv Z_{nn}$ from
the equation (\ref{32B}) with the anomalous dimension $\gamma_{n}=
\widetilde{\cal D}_{\mu} \ln Z_{n}$.

Finally, consider irreducible $l$-th rank operators of the form
\begin{equation}
F_{nl}= \partial_{i_{1}}\theta\cdots\partial_{i_{l}}\theta\,
(\partial_{i}\theta\partial_{i}\theta)^{p}+\cdots, \qquad n=l+2p
\label{Fnl}
\end{equation}
with $d_{F}=0$, $d_{F}^{\omega}=-n/2$ (note that $F_{n0}=F_{p}$). Here the
dots stand for the appropriate subtractions involving the Kronecker
$\delta$ symbols, which ensure that the resulting expressions are traceless
with respect to any given pair of indices, for example, $\partial_{i}\theta
\partial_{j}\theta - \delta_{ij}\partial_{k}\theta\partial_{k}\theta /d$.
Of course, the numbers $n$ and $l$ have the same parity, that is, they can
only be simultaneously even or odd. Like for the operators (\ref{Fn}), one
can show that the operators (\ref{Fnl}) mix only with each other in
renormalization, the corresponding renormalization matrix is triangular,
the critical dimensions are determined by its diagonal elements $Z_{nl}$,
and the anomalous dimensions are
$\gamma_{nl}= \widetilde{\cal D}_{\mu} \ln Z_{nl}$.

One important remark is relevant here. The matrix elements $Z_{n}$
for the operators $F_{n}$ and $Z_{nl}$ for $F_{nl}$ are determined
by the 1-irreducible correlation functions $\langle F
\theta\cdots\theta \rangle_{\rm 1-ir}$, in which the number of the
fields $\theta$ equals to their number in the operator $F$. It is
easily seen that the corresponding Feynman diagrams do not involve
the bare propagator $\langle \theta\theta\rangle_{0}$ from
(\ref{lines}), and, hence, the correlation function of the scalar
random noise (\ref{2}). As a result, the critical dimensions of
the operators (\ref{Fn}) and (\ref{Fnl}) are completely
independent of the form of the scalar forcing in Eq. (\ref{1}). We
also note that for the same reason the operators (\ref{Fnl}) with
equal $n$ and different $l$ do not mix in renormalization: this is
forbidden by SO$(d)$ symmetry, which is present in the relevant
diagrams even if the correlation function (\ref{2}) is
anisotropic.

In contrast to (\ref{Deltas}), the critical dimensions $\Delta_{n}$ and
$\Delta_{nl}$ of the operators (\ref{Fn}) and (\ref{Fnl}) are nontrivial;
they are calculated as series in $\eps$:
\begin{equation}
\Delta_{nl} = \sum_{k=1}^{\infty} \eps^{k} \, \Delta_{nl}^{(k)}
\label{answer}
\end{equation}
(of course, $\Delta_{p}=\Delta_{n0}$ with $n=2p$). An important
exception is provided by the dimension of the operator $F_{1}=
(\partial_{i}\theta \partial_{i}\theta)$, the local dissipation
rate of the scalar fluctuations, which can be found exactly; see
below. The calculation to the order $\eps^{2}$ (two-loop
approximation) will be presented in Sec.~\ref{sec:Dimensions} in
detail, and here we give only the first-order result:
\begin{equation}
\Delta_{nl}^{(1)} =   -\frac{n\,(n-2)}{3(d+2)} + \frac{(d+1)\,l\,
(d+l-2)} {3(d-1)(d+2)}.
\label{Qnp}
\end{equation}
Expression (\ref{Qnp}) was already presented in \cite{RG3}. It
differs only by an overall factor from its analog for Kraichnan's
model \cite{GK,RG3} or Gaussian model with finite correlation time
\cite{RG3}.\footnote{More precisely, the first-order result for
$\Delta_{nl}$ in Kraichnan's model is obtained from
(\protect\ref{answer}), (\protect\ref{Qnp}) after the substitution
$\eps\to 3\xi/2$, where $\xi$ is the exponent in the
velocity--velocity correlation function $\langle vv \rangle
\propto \delta(t-t')/k^{d+\xi}$. Thus for the ``physical'' values
($\eps=2$ and $\xi=4/3$) they coincide.}

The result $\Delta_{1}=\Delta_{20}=0$ in (\ref{Qnp}) is in fact valid to
all orders in $\eps$. This can be demonstrated using the Schwinger
equation of the form
\begin{equation}
\int{\cal D}\Phi {\delta} \left[ \theta(x) \exp S_{R}( \Phi)
+ A \Phi\right]/{\delta\theta'(x)}  =0,
\label{Schwi}
\end{equation}
where $S_{R}$ is the renormalized action (\ref{Rac}) and the notation
introduced in (\ref{14}) is used. (We recall that in the general sense of
the term, Schwinger equations are any relations stating that any
functional integral of a total variational derivative vanishes;
see e.g. Ref. \cite{Book1}.) Equation (\ref{Schwi}) can be rewritten
in the form
\begin{equation}
\langle\langle \theta' D_{\theta} \theta - \nabla_{t}[\theta^{2}/2] +
\nu Z_{u}Z_{\nu}\partial^{2}[\theta^{2}/2]-\nu Z_{u}Z_{\nu} F_{1}\rangle\rangle
_{A}=-A_{\theta'} \delta W_{R}(A)/\delta A_{\theta},
\label{Schwi2}
\end{equation}
where $D_{\theta}$ is the correlation function (\ref{2}),
$\langle\langle \dots\rangle\rangle _{A}$ denotes the averaging with the
weight $ \exp [S_{R}(\Phi) + A \Phi]$, $W_{R}$ is determined by the Eq.
(\ref{14}) with the replacement $S\to S_{R}$, and the argument $x$ common
to all the quantities is omitted.

The quantity $\langle\langle F\rangle\rangle _{A}$ is the generating
functional of the correlation functions with one operator $F$ and any number
of the fields $\Phi$, therefore the UV finiteness of the operator $F$ is
equivalent to the finiteness of the functional
$\langle\langle F\rangle\rangle _{A}$. The quantity on the right-hand side
of (\ref{Schwi2}) is UV finite (a derivative of the renormalized functional
with respect to finite argument), and so is the operator in its left-hand
side. Our operator $F_{1}$ does not admix in renormalization to the operator
$\theta' D_{\theta}\theta$ ($F_{1}$ contains too many fields $\theta$) and
to the operators $\nabla_{t}[\theta^{2}/2]$ and $\partial^{2}[\theta^{2}/2]$
(they have the form of total derivatives, and $F_{1}$ does not reduce to
such form). On the other hand, the operator
$\theta' D_{\theta}\theta$ does not admix to $F_{1}$ (it is
nonlocal, and $F_{1}$ is local), while the derivatives
$\nabla_{t}[\theta^{2}/2]$ and $\partial^{2}[\theta^{2}/2]$
do not admix to $F_{1}$ owing to the fact that each field
$\theta $ enters in the counterterms of the operators
$F_{n}$ only in the form of derivative $\partial\theta$
(see above). Therefore, all three types of operators entering
into the left-hand side of Eq. (\ref{Schwi2}) are independent,
and they must be UV finite separately.

Since the operator $\nu Z_{u}Z_{\nu} F_{1}$ is UV finite, it coincides
with its finite part, i.e., $\nu Z_{u}Z_{\nu}F_{1}=\nu F_{1}^{R}$,
which along with the relation $F_{1}=Z_{1}F_{1}^{R}$ gives
$Z_{1}^{-1}=Z_{u}Z_{\nu}$ and therefore $\gamma_{1}
=-\gamma_{u}-\gamma_{\nu}$. We recall that at the fixed point
$\gamma_{\nu}^{*}=2\eps/3$ and $\gamma_{u}^{*}=0$ (the latter
equality follows from the relation $\beta_{u}=-u\gamma_{u}$ and $u_{*}>0$),
so that $\gamma_{1}^{*}=-2\eps/3$. From the relations (\ref{32B}) and
Table \ref{table1} one obtains $\Delta_{n}=2n\eps/3 + \gamma_{n}^{*}$.
Combining these expressions gives the desired exact result $\Delta_{1}=0$.
It will be used later to prove the absence of anomaly for the second-order
structure function. What is more, it can be used to essentially simplify the
calculation of the other dimensions $\Delta_{nl}$ to order $\eps^{2}$;
see Sec.~\ref{sec:Dimensions}.

\section{Operator product expansion and composite operators}
\label{sec:OPE}

The representation (\ref{RGR}) describes the behavior of the
correlation function $G(r)$ for $\Lambda r \gg1$ and any fixed
value of $mr$. In particular, for the structure functions
(\ref{struc}) using the data from Table \ref{table1},  Eq.
(\ref{Deltas}) and the definitions $g_{0} = D_{0}/\nu_0^3 =
\Lambda^{2\eps}$ one obtains
\begin{equation}
S_{n}({\bf r}) = D_{0}^{-n/2}\, r^{n\Delta_{n}}\, \xi_n (mr).
\label{strucS}
\end{equation}
For our model, odd structure functions vanish, but they become
nontrivial if, for example, the random force in Eq. (\ref{1}) is
replaced by an imposed constant gradient. The inertial range
corresponds to the additional condition that $mr\ll 1$. The form
of the function $\xi(mr)$ is not determined by the RG equations
themselves; in the theory of critical phenomena, its behavior for
$mr\to0$ is studied using the well-known Wilson operator product
expansion; see e.g. Refs.~\cite{Zinn,Book3}. This technique is
also applicable in the theory of turbulence
\cite{Book3,JETP,UFN,turbo}.

According to the OPE, the equal-time random quantity in the left-hand
side of Eq. (\ref{struc}) at ${\bf x}\equiv ({\bf x}_{1} + {\bf x}_{2} )/2
= {\const}$ and ${\bf r}\equiv {\bf x}_{1} - {\bf x}_{2}\to 0$ can be
represented in the form
\begin{equation}
[\theta(t,{\bf x}_{1})-\theta(t,{\bf x}_{2})]^{n}
=\sum_{F} C_{F} ({\bf r}) F(t,{\bf x}),
\label{OPE}
\end{equation}
where the functions $C_{F}$ are the Wilson coefficients regular in $m^{2}$
and $F$ are various composite operators (more precisely, see below).

In general, the operators $F$ entering the expansion like (\ref{OPE}) are
all possible renormalized local operators, allowed by the symmetry of the
model and the quantity in the left-hand side. In practice, they can be
found as the monomials which appear in the corresponding Taylor expansions,
and all possible operators that admix to them in renormalization. If
these operators have additional vector indices, they are contracted with
the corresponding indices of the coefficients $C_{F}$.

With no loss of generality it can be assumed that the expansion in Eq.
(\ref{OPE}) is made in the operators with definite critical dimensions
$\Delta_{F}$. The structure function (\ref{struc}) in renormalized
variables is obtained by averaging Eq. (\ref{OPE}) with the weight
$\exp S_{R}$ with $S_{R}$ from Eq. (\ref{Rac}), then the quantities
$\langle F \rangle$ appear on the right hand side. Their asymptotic
behavior for $m\to0$ is found from the corresponding RG equations and
has the form $\langle F \rangle \propto  m^{\Delta_{F}}$.

Combining the operator product expansion (\ref{OPE}) with the asymptotic
representation (\ref{strucS}) we therefore find the following expression
for the scaling functions $\xi_{n}(mr)$ in the region $mr\ll1$:
\begin{equation}
\xi_{n}(mr)=\sum_{F}A_{nF}\,(mr)^{\Delta_{F}},
\label{OR}
\end{equation}
with the coefficients $A_{nF}(mr)$ regular in $(mr)^{2}$.

Obviously, the leading term of the asymptotic behavior of the function
(\ref{OR}) for $(mr)\ll1$ is determined by the operator with the minimal
dimension $\Delta_{F}$.
The following additional considerations should be
taken into account.

(i) With no loss of generality, it can be assumed that the expansion
(\ref{OPE}) is made in irreducible traceless tensor composite operators.
In the isotropic case, the mean values of all nonscalar irreducible
operators vanish, and their dimensions do not appear in the right-hand side
of Eq. (\ref{OR}).

(ii) Owing to Galilean invariance of the model and the structure functions
(\ref{struc}), only invariant operators appear in the expansion (\ref{OPE}).

(iii) The action functional (\ref{action}) and the functions (\ref{struc})
are invariant with respect to the shift $\theta\to\theta+\const$, and the
operators on the right-hand side of (\ref{OPE}) must also be invariant.
This means that they can involve the fields $\theta$ only in the form of
(covariant) derivatives $\partial_{i}\theta$ or $\nabla_{t}\theta$.

(iv) Using the linearity of the equation (\ref{1}) in the field $\theta$,
one can show that for any operator $F$ that appear in the expansion like
(\ref{OPE}), the number of the fields $\theta$ cannot exceed their
total number on the left-hand side.

Finally, we recall that $\Delta_{F}=d_{F}+O(\eps)$. Thus we may conclude
that, at least for small $\eps$, the leading terms in the small-$(mr)$
behavior for the even function $S_{2n}$ is given by one of the operators
$F_{k}$ from (\ref{Fn}) with $k\le n$. Indeed, any additional derivative or a
field different from $\theta$ leads to an increase of the dimension
$\Delta_{F}$; the operators $F_{k}$ with $k>n$ are forbidden by the property
(iv), while the operators containing more fields than derivatives are
forbidden by (iii). From the explicit form (\ref{Qnp}) it follows that
the dimension $\Delta_{k}$ monotonically decreases as $k$ grows. We finally
conclude that the leading term in (\ref{OR}) is given by the contribution
of the operator $F_{n}$ from (\ref{Fn}) and
\begin{equation}
S_{2n} = D_{0}^{-n} r^{-2n\Delta_{\theta}} (mr)^{\Delta_{n}}
\propto r^{n+\gamma_{n}^{*}}
\label{Anomal}
\end{equation}
with the dimension $\gamma_{n}^{*}$ defined above Eq. (\ref{Fnl}).

This establishes the existence of anomalous scaling for the
passive scalar field in our model with the identification
$\zeta_{2n}=-2n\Delta_{\theta}+ \Delta_{n}=n+\gamma_{n}^{*}$; see
text below Eq.~(\ref{struc}).

If the large-scale anisotropy is introduced to the problem by the correlation
function of the scalar noise (\ref{2}), tensor operators acquire nonzero
mean values and their dimensions also appear on the right-hand side of the
expansion (\ref{OR}). An $l$-th rank irreducible operator gives rise to a
term in $\xi(mr)$ proportional to the spherical harmonics $Y_{lm}$ for $d=3$
or their analogs for general $d$. In the special case of uniaxial anisotropy,
when the function $C$ in Eq. (\ref{2}) depends on a fixed unit vector
${\bf n}$ in addition to ${\bf r}$, they reduce to terms proportional
to the Gegenbauer polynomials $P_{l}$ (Legendre polynomials for $d=3$).

Repeating the above analysis we conclude that the leading term in the
$l$-th anisotropic sector of the scaling function $\xi(mr)$ for $mr\ll1$
is determined by the $l$-th rank tensor operator $F_{nl}$ (\ref{Fnl})
with the dimension $\Delta_{nl}$ from (\ref{Qnp}). In particular, for the
case of uniaxial anisotropy
\begin{equation}
S_{n} \propto \ \cdots + P_{l}(\cos\vartheta)\,
r^{n+\gamma_{nl}^{*}} + \cdots,
\label{Animal}
\end{equation}
where $\gamma_{nl}^{*}$ is defined in Sec.~\ref{sec:Operators}
below Eq. (\ref{Fnl}), $\vartheta$ is the angle between the
directions ${\bf n}$ and ${\bf r}$, and the dots stand for the
contributions of the other anisotropic sectors. It remains to note
that the odd functions $S_{2n+1}$ are nontrivial if, for example,
the random force in Eq. (\ref{1}) is replaced by an imposed
constant gradient, and their leading terms are then determined by
the vector operators $F_{2n+1,1}$.

\section{Calculation of the critical dimensions of operators
$F_{\lowercase{nl}}$ in the two-loop approximation}
\label{sec:Dimensions}

\subsection{General scheme and the relevant diagrams} \label{sec:General}

From now on, we shall consider composite operators (\ref{Fnl}) in
the model without the scalar noise in Eq. (\ref{1}), that is, with
$D_{\theta}=0$ in the action functional (\ref{action}). The
stirring force in Eq. (\ref{1.1}), that is, the term with $D_{v}$
in the action functional (\ref{Rac}), should be retained. Then the
operators (\ref{Fnl}) are renormalized multiplicatively,
$F_{nl}=Z_{nl}F_{nl}^{R}$; see Sec. \ref{sec:Operators}. The
renormalization constants $Z_{nl}=Z_{nl}(g,u,d,\eps)$ are
determined by the requirement that the 1-irreducible correlation
function
\begin{equation}
\bigl\langle F_{nl}^{R} (x) \theta(x_{1})\dots\theta(x_{n})
\bigr\rangle_{\rm 1-ir}= Z_{nl}^{-1}\bigl\langle
F_{nl}(x) \theta(x_{1})\dots\theta(x_{n})\bigr\rangle
_{\rm 1-ir} \equiv Z_{nl}^{-1}\Gamma_{nl} (x;x_{1},\dots, x_{n})
\label{req}
\end{equation}
be UV finite in renormalized theory (\ref{Rac}), (\ref{RaV}), i.e., have
no poles in $\eps$ when expressed in renormalized variables
(\ref{renorm}). This is equivalent to the UV finiteness of the product
$Z_{nl}^{-1}\Gamma_{nl}(x;\theta)$, in which
\begin{equation}
\Gamma_{nl} (x;\theta) = \frac{1}{n!}\, \int d x_{1}\dots \int d x_{n}\,
\Gamma_{nl} (x;x_{1},\dots, x_{n})\, \theta(x_{1})\dots\theta(x_{n})
\label{gena}
\end{equation}
is a functional of the field $\theta(x)$. In the zeroth
approximation, the functional (\ref{gena}) coincides with the
operator $F_{nl}(x)$, and in higher orders the kernel $\Gamma_{nl}
(x;x_{1},\dots, x_{n})$ is given by the sum of diagrams shown in
Fig. \ref{table2} up to the two-loop order with their symmetry
coefficients (coefficients of the diagrams Nos 2, 4 and 6 are
equal to unity, while coefficients of the diagrams Nos 8 and 9 are
nontrivial, but they are not shown because we will not need them;
see below). The dashed lines denote the propagators
(\ref{linesV}), while the solid lines denote the propagators
(\ref{lines}); the slashed ends correspond to the auxiliary fields
${\bf v}'$ and $\theta'$, the ends without a slash correspond to
the fields ${\bf v}$ and $\theta$. Since we are working with the
renormalized theory, the replacements $\nu_0\to\nu$, $\kappa_0\to
u\nu$ should be made in the denominators of Eqs. (\ref{linesV}),
(\ref{lines}), and the amplitude in $\langle vv\rangle_{0}$ should
be expressed in renormalized variables: $D_{0}=g\mu^{2\eps}\nu^3$
(see text below Eq. (\ref{renorm})). The diagrams Nos 1--7 involve
only the vertex (\ref{vertex}) while the diagrams Nos 8 and 9 also
involve the vertex (\ref{vertexV}). One dashed line attached to
any of the vertices (\ref{vertexV}) must be slashed; there are two
variants for the diagram No 8 and three variants for No 9. We do
not show these variants explicitly (and do not show the slashes
and symmetry coefficients for these diagrams), because, as we
shall see below, the total contribution of the diagrams Nos 2, 6,
8, and 9 can be found without the practical calculation of their
individual contributions.

The thick dots in the diagrams correspond to the vertices of the
composite operator $F_{nl}$ from (\ref{Fnl}). According to the
general rules of the universal diagrammatic technique (see e.g.
Ref. \cite{Book1}), for any composite operator $F(x)$ built of the
fields $\theta$, the vertex with $k\ge0$ attached lines
corresponds to the vertex factor
\begin{equation}
V_{k} (x;\, x_{1}, \dots, x_{k}) \equiv \delta^{k}
F(x) / {\delta\theta(x_{1}) \dots\delta\theta(x_{k})}.
\label{BigV}
\end{equation}
The arguments $x_{1}\dots x_{k}$ of the quantity (\ref{BigV}) are
contracted with the arguments of the upper ends of the lines
$\langle\theta\theta' \rangle_{0}$ attached to the vertex. For our
operators (\ref{Fnl}), built solely of the gradients
$w_{i}(x)\equiv
\partial \theta(x)/\partial x_{i}$ at a single spacetime point
$x$, the factors (\ref{BigV}) contain the product $\partial_{i_1}
\delta(x-x_{1}) \dots \partial_{i_k} \delta(x-x_{k})$, and the
integrations over $x_{1}\dots x_{k}$ are easily performed: the
derivatives move onto the upper ends of the corresponding lines
$\langle\theta\theta' \rangle_{0}$ attached to the vertex, and
their arguments $x_{1}\dots x_{k}$ are substituted with $x$. After
the derivatives have been moved inside the diagram, the remaining
vertex factor for the operator $F(x)$ can be understood as a usual
derivative:
\begin{equation}
V_{i_{1}\dots i_{k}} (x) = \partial^{k} F(x) /
\partial w_{i_{1}} (x)\dots \partial w_{i_{k}} (x) .
\label{99}
\end{equation}

The analysis of the diagrams shows that for any argument $x_{s}$ in the
quantity (\ref{gena}), the corresponding spatial derivative is isolated
as an external factor from each diagram. Using the integration by parts,
these derivatives can be moved onto the corresponding fields
$\theta(x_{s})$, so that the quantity (\ref{gena}) can be represented
as a functional of the vector field $w_{i}=\partial_{i}\theta$:
\begin{equation}
\Gamma_{nl} (x;\theta) = \frac{1}{n!}\, \int d x_{1}\dots \int d x_{n}\,
\widetilde \Gamma_{nl}^{i_{1}\dots i_{n}} (x;x_{1},\dots, x_{n})\,
w_{i_{1}}(x_{1}) \dots w_{i_{n}}(x_{n}).
\label{gena2}
\end{equation}

The diagrams that determine the kernel $\widetilde \Gamma$ in (\ref{gena2})
contain only logarithmic UV divergencies. Therefore, in order to find the
constant $Z_{nl}^{-1}$ it is sufficient to calculate the functional
$\widetilde \Gamma$ with some special choice of its functional argument
$w_{i}$, namely, one can replace it by its value at the fixed
point $x$, the argument of the operator $F_{nl}$ in Eqs. (\ref{req}). Now
the product $w_{i_{1}}(x)\dots w_{i_{n}}(x)$ can be taken outside
the integrals over $x_{1},\dots, x_{n}$ in (\ref{gena2}), so that the
functional $\Gamma_{nl} (x;\theta)$ turns to a local composite operator. The
integration of the remaining function $\widetilde \Gamma_{nl}$ over
$x_{1},\dots, x_{n}$ gives a quantity independent of any coordinate variables,
and its vector indices can only be those of Kronecker delta symbols. Their
contraction with the indices of the product $w_{i_{1}}(x)\dots w_{i_{n}}(x)$
gives rise to the original operator $F_{nl}(x)$ with some scalar coefficient
$\overline\Gamma$. The integration over $x_{1},\dots, x_{n}$ means that in
the Fourier representation, the corresponding correlation function is
calculated with all its momenta set equal to zero, which is always implied
in what follows. The IR regularization is provided by the parameter $m$ in
the function (\ref{1.2}). In a compact notation one can write:
\begin{equation}
Z_{nl} \equiv Z_{F}, \qquad F_{nl}(x) \equiv F, \qquad
\Gamma_{nl}(x;\theta) \equiv \Gamma = F \overline\Gamma.
\label{brev}
\end{equation}

In the MS scheme all renormalization constants have the form
\begin{eqnarray}
Z_{F}= 1+\sum _{k=1}^{\infty } Z_{F}^{(k)} \eps ^{-k},
\label{1.30}
\end{eqnarray}
where the coefficients $Z_{F}^{(k)}=Z_{F}^{(k)}(g,u,d)$ are
independent of $\eps$. Then for the corresponding anomalous
dimension $\gamma_{F}$ from the definitions (\ref{RGF}) and
relations (\ref{RGF2}) for the $\beta$ functions one obtains
$\gamma_{F}\equiv \widetilde{\cal D}_{\mu} \ln Z_{F} =
[\beta_{g}\partial_{g}+\beta_{u}\partial_{u}] \ln Z_{F} = -2 {\cal
D}_{g} Z_{F}^{(1)} +$ terms containing poles in $\eps$ (we recall
that ${\cal D}_{g}=g\partial_{g}$). However, the pole parts must
cancel each other owing to the UV finiteness of the anomalous
dimension $\gamma_{F}$. We therefore arrive at the expression
\begin{eqnarray}
\gamma_{F} = -2 {\cal D}_{g} Z_{F}^{(1)},
\label{130}
\end{eqnarray}
that is, in order to find the dimension $\gamma_{F}$ it is
sufficient to find the first-order residue $Z_{F}^{(1)}$ in the
expansion (\ref{1.30}). If desired, the higher-order residues
$Z_{F}^{(k)}$ with $k\ge2$ can be calculated to check the
aforementioned cancellation (and thus the correctness of the
calculations).

Now we turn to the practical calculation of the diagrams needed to
determine the coefficients $Z_{nl}^{(1)}$ in the constants $Z_{nl}$
from (\ref{req})  to order $g^{2}$ (two-loop approximation).

\subsection{Scalarization of the diagrams and contractions of basic
tensor structures} \label{sec:Scal}

The contribution of a specific diagram into the functional $\Gamma$ in
(\ref{gena2}) for any composite operator $F$, built of the gradients
$w_{i}=\partial_{i} \theta$, is represented in the form
\begin{equation}
\Gamma = V_{12\dots} \, I^{ab\dots}_{12\dots} \, w_{a} w_{b} \dots ,
\label{V1}
\end{equation}
where $V_{12\dots}$ is the vertex factor (\ref{99}),
$I^{ab\dots}_{12\dots}$ is the ``internal block''  of the diagram
with free indices, the product $w_{a} w_{b} \dots$ corresponds to
external lines. The numerical indices $1,2,\dots$ will always be
understood as $i_{1}, i_{2}, \dots$, their number in Eq.
(\ref{V1}) equals to the number of the letter indices $a,b,\dots$
and is determined by the number of ``rays,''  that is, the number
$k$ of lines that attach to the vertex of the operator. These
lines are given by products of the propagators
$\langle\theta\theta'\rangle_0$ from Eq. (\ref{lines}) and are
connected by the lines $\langle vv\rangle_0$ and $\langle
vv'\rangle_0$ from Eq. (\ref{linesV}); see examples in Fig.
\ref{table2}. For the two-loop calculation, it is sufficient to
consider diagrams with $k=2$ and 3, because the diagram No 5 with
$k=4$ factorizes into products of the blocks with $k=2$ and,
therefore, gives no contribution to the first-order residue
$Z_{F}^{(1)}$. [This is true only if the IR regularization in the
correlator (\protect\ref{1.2}) is provided by the sharp cutoff at
$k=m\equiv1/\ell$, so that the one-loop integral is a pure pole in
$\eps$; see Eq. (\protect\ref{oneloop2}) below. If the
regularization is provided, e.g., by the substitution $k^{2}\to
k^{2}+m^{2}$ in (\protect\ref{1.2}), the one-loop integral
contains an $O(\eps^{0})$ term, the diagram No 5 contains a
first-order pole in $\eps$ and contributes to $Z_{F}^{(1)}$.
However, the total value of $Z_{F}^{(1)}$ in the MS scheme is
independent of the choice of IR regularization.]

Since the vertex factor (\ref{99}) and the product $w_{a} w_{b}
\dots$ are symmetrical with respect to any permutations of their
indices, the quantity $I^{ab\dots}_{12\dots}$ in Eq.~(\ref{V1}) is
automatically symmetrized with respect to any permutations of the
letter indices $a,b,\dots$ and the numerical indices $1,2,\dots$.
In what follows, such symmetrization will be denoted by the symbol
${\cal S}$. For any fixed number of rays $k$, the quantity ${\cal
S}\, I$ is represented as a linear combination
\begin{equation}
{\cal S}\, I = \sum_{i} B_{i}\, S_{i}
\label{V4}
\end{equation}
of certain basis tensor structures $S_{i} = (S_{i})^{ab\dots}_{12\dots}$
with certain numerical coefficients $B_{i}$. There are two such structures
for the $k$-ray diagrams with $k=2$ and 3; they have the forms:
\begin{mathletters}
\label{V5}
\begin{eqnarray}
k&=&2: \qquad S_{1} = \Sym\, [\delta_{1a}\delta_{2b}], \qquad
              S_{2} = \Sym\, [\delta_{12}\delta_{ab}];
\label{V5a} \\
k&=&3: \qquad S_{1} = \Sym\, [\delta_{1a}\delta_{2b}\delta_{3c}], \qquad
S_{2} = \Sym\, [\delta_{12}\delta_{ab}\delta_{3c}].
\label{V5b}
\end{eqnarray}
\end{mathletters}

The quantities which will be directly calculated from the diagrams are not
the coefficients $B_{i}$ themselves, but the following scalar quantities
related to them:
\begin{equation}
A_{i} = {\rm tr}\, \bigl[(S_{i})^{ab\dots}_{12\dots} \, \Sym\,
I^{ab\dots}_{12\dots}\bigr] = {\rm tr}\, [S_{i} \cdot \Sym\, I\,],
\label{V6}
\end{equation}
where the symbol tr denotes the contraction with respect to all repeated
indices, which will not be shown explicitly. It is therefore necessary to
express the coefficients $B_{i}$ in (\ref{V4}) in terms of the quantities
(\ref{V6}). We omit the derivation, which is identical to the case of
Kraichnan model (see Ref. \cite{cube} for detail) and give only the answer:
\begin{mathletters}
\label{V10}
\begin{eqnarray}
k&=&2: \quad B_{1} = 2\alpha_{2} [dA_{1}-A_{2}], \quad
B_{2} = \alpha_{2} [-2A_{1}+(d+1)A_{2}] \quad {\rm with} \
\alpha_{2} \equiv [(d-1)d(d+2)]^{-1};
\label{V10a}
 \\
k&=&3: \quad B_{1} = 6\alpha_{3} [(d+2)A_{1}-3A_{2}],
\quad  B_{2} = 9\alpha_{3} [-2A_{1}+(d+1)A_{2}]
\quad {\rm with} \ \alpha_{3} \equiv [(d-1)d(d+2)(d+4)]^{-1}.
\label{V10b}
\end{eqnarray}
\end{mathletters}

The next step is the contraction of the quantities
$I^{ab\dots}_{12\dots}$ in Eq. (\ref{V1}) with external factors:
the vertex factor $V_{12\dots}$ of the composite operator and the
product $w_{a}w_{b}\dots$. Again, the derivation is identical to
the case of Kraichnan model (see \cite{cube} for details) and we
only give the result:
\begin{equation}
\Gamma = F \overline\Gamma, \qquad  \overline\Gamma =\sum_{i} k_{i} B_{i},
\label{V21}
\end{equation}
where we use the notation (\ref{brev}) and the coefficients $k_{i}$ have
the forms:
\begin{mathletters}
\label{V22}
\begin{eqnarray}
k&=&2: \qquad k_{1} =n(n-1), \quad k_{2}= \lambda _{nl}, \quad
{\rm where} \quad \lambda _{nl}= (n-l)(d+n+l-2);
\label{V22a}  \\
k&=&3: \qquad k_{1}=n(n-1)(n-2), \quad k_{2}= (n-2)\lambda _{nl}.
\label{V22b}
\end{eqnarray}
\end{mathletters}

Finally, combining Eqs. (\ref{V10}) and (\ref{V22}) we express the function
$\Gamma$ in the scalar quantities $A_{i}$:
\begin{equation}
\overline\Gamma =\sum_{i} p_{i} A_{i},
\label{V21A}
\end{equation}
where
\begin{mathletters}
\label{PA}
\begin{eqnarray}
k&=&2: \qquad p_{1} = 2\alpha_{2} \left[n(n-2)(d-1)+\lambda_{l} \right],
\quad
p_{2} = \alpha_{2} \left[n(n+d)(d-1)-(d+1)\lambda_{l} \right];
\label{PAa}  \\
k&=&3: \qquad p_{1} = 6\alpha_{3}(n-2) \left[ n(n-4)(d-1)+3
\lambda_{l} \right], \quad
p_{2} = 9\alpha_{3}(n-2) \left[n(n+d)(d-1)-(d+1)\lambda_{l} \right]
\label{PAb}
\end{eqnarray}
\end{mathletters}
with $A_{1,2}$ from (\ref{V6}), $\alpha_{2,3}$ from (\ref{V10}), and
$\lambda_{l}=l(l+d-2)$. Note that the coefficients $p_{i}$ in (\ref{PAb})
vanish for $n=2$ (in general, the diagrams with $k$ rays give no
contribution to the functions $\Gamma$ for the operators $F_{nl}$
with $n<k$). Also note that the coefficient $p_{1}$ in (\ref{PAa}) vanishes
for $n=2$, $l=0$; this fact will be important in what follows.

\subsection{General relations for the anomalous dimensions} \label{sec:Anal}

Let us denote by ${R} \equiv Z_{F}^{(1)}$ the first-order
coefficient in the expansion (\ref{1.30}) for the renormalization
constant $Z_{F}$. In the perturbation theory, it is calculated as
the series $R=\sum_{s=1}^{\infty} g^{s}R_{s}$ in powers of the
renormalized coupling constant $g$, with coefficients $R_{s}$
depending on $u$ and $d$. In their turn, they can be written as
the sums $R_{s}=\sum_{k=2} R_{s}^{(k)}$, where $R_{s}^{(k)}$ is
the total contribution of the diagrams with $k$ rays. As
illustrated by Fig. \ref{table2}, the number of terms in the
latter sum is always finite: for $s=1$, there is the only
contribution with $k=2$, for $s=2$, there are contributions with
$k=2$ and 3, and so on. Similar decompositions can be written for
the coefficients in front of the $1/\eps$ contributions to the
scalar quantities $A_{i}$ in (\ref{V6}). The corresponding
coefficients will be denoted by $a_{is}^{(k)}$, where $i=1$, 2 is
the number of structure, $s$ is the order in $g$, and $k$ is the
number of rays. From the definition of the constants $Z_{F}$ and
expressions (\ref{brev}) and (\ref{V1}) it follows that, at the
same time, $a_{is}^{(k)}$ is the total contribution of the $i$-th
structure to the quantity $R_{s}^{(k)}$, that is,
$R_{s}^{(k)}=\sum_{i} p_i^{(k)} a_{is}^{(k)}$. Here $p_i^{(k)}$
are the coefficients (\ref{PA}) in which the numbers of rays are
explicitly shown.

We shall calculate the first $R_{1}$ and the second $R_{2}$ terms
(two-loop approximation), which involves diagrams with 2 and 3
rays. Using the above definitions, expressions (\ref{130}) and
(\ref{PAa}), and neglecting the terms of order $g^{3}$ and higher,
we can write the following representation to the two-ray
contribution $\gamma^{(2)}_{F}$ to the anomalous dimension
$\gamma_{F}\equiv \gamma_{nl}$ at the fixed point (\ref{points}):
\begin{mathletters}
\label{Atas}
\begin{equation}
\gamma^{(2)}_{F}= -2 \left\{ p_{1}\left[g_{*}a_{11}^{(2)}+2g_{*}^{2}
a_{12}^{(2)} \right]  + p_{2} \left[g_{*}a_{21}^{(2)}+2g_{*}^{2}
a_{22}^{(2)} \right] \right\}
\label{Atas1}
\end{equation}
with the coefficients $p_{1,2}$ from (\ref{PAa}). Similarly, for the
three-ray contribution one can write
\begin{equation}
\gamma^{(3)}_{F}= -4 g_{*}^{2} \left\{ p_{1} a_{12}^{(3)} + p_{2}
 a_{22}^{(3)} \right\}
\label{Atas2}
\end{equation}
\end{mathletters}
with $p_{1,2}$ from (\ref{PAb}). Since diagrams with three rays appear only
in order $g^{2}$, the contribution of order $g$ in the latter expression is
absent. We recall that the quantities $a_{is}^{(k)}$ depend on $u$ and $d$.
In (\ref{Atas}), the substitution $u=u_{*}$ is implied.

The quantities (\ref{Atas}) should be calculated up to the order
$\eps^{2}$. The contribution $\gamma^{(3)}_{F}$ is of order
$g^{2}$, so that in (\ref{Atas2}) it is sufficient to take the
coordinates $g_{*}$, $u_{*}$ of the fixed point in the
lowest-order approximation: $g_{*}=g_{*}^{(1)}\eps$,
$u_{*}=u_{*}^{(0)}$ with $g_{*}^{(1)}$, $u_{*}^{(0)}$ from
(\ref{FP1}). [We recall that the upper indices for $g_{*}$ and
$u_{*}$ denote the orders of the expansion in $\eps$; see Eq.
(\ref{FPE}).] The contribution $\gamma^{(2)}_{F}$ contains terms
of order $g$ and $g^{2}$. Therefore, in (\ref{Atas1}) one should
take into account the leading correction terms to the coordinates
of the fixed point, denoted as $g_{*}^{(2)}$ and $u_{*}^{(1)}$ in
(\ref{FPE}). We are going to show, however, that the quantity
(\ref{Atas1}) can in fact be calculated without knowledge of the
coefficients $g_{*}^{(2)}$, $u_{*}^{(1)}$, $a_{21}^{(2)}$ and
$a_{22}^{(2)}$.

We recall that the dimension $\gamma_{nl}^{*}$ for $n=2$, $l=0$ is known
exactly: $\gamma_{1}^{*}\equiv \gamma_{20}^{*}=-2\eps/3$; see the end of
Sec.~\ref{sec:Operators}. This dimension is completely determined by the
two-ray contribution $\gamma^{(2)}_{F}$ from (\ref{Atas1}), while
$\gamma^{(3)}_{F}$ for $n=2$ vanishes. The coefficient $p_{1}$ in
(\ref{PAa}) for $n=2$, $l=0$ vanishes, while $p_{2}$ remains nontrivial:
$p_{2}=2/d$. Thus from Eq. (\ref{Atas1}) we immediately find the
following exact answer for the quantity in the second square brackets:
\begin{equation}
\left[g_{*}a_{21}^{(2)}+2g_{*}^{2} a_{22}^{(2)} \right] = d\eps/6.
\label{Exact}
\end{equation}
This means that the only contribution that survives in the
left-hand side is $\eps g_{*}^{(1)} a_{21}^{(2)}|_{u=u_*^{(0)}}$,
while the $O(\eps^{2})$ contributions with $g_{*}^{(2)}$,
$u_{*}^{(1)}$ and $a_{22}^{(2)}$ must cancel each other. In order
to verify the validity of our calculations, we checked this
cancellation for $d=3$. All the dependence on $n$ and $l$ in Eqs.
(\ref{Atas}) comes from the coefficients $p_{1,2}$, so that the
expression (\ref{Exact}) determines the contribution in the second
square brackets in (\ref{Atas1}) for all $n$ and $l$.

Since for $n=2$ and $l=0$ the coefficient $p_{1}$ vanishes, the
exact answer for $\gamma_{20}^{*}$ gives no information about the
quantity in the first square brackets. However, from the explicit
expression for the only one-loop diagram in Fig. \ref{table2} it
is not difficult to see that the corresponding structure $A_1$
vanishes identically, so that $a_{11}^{(2)}=0$ in (\ref{Atas1}).
Indeed, in the quantity $A_1$ the upper (letter) indices of the
tensor $I^{ab\dots}_{12\dots}$ in Eq.~(\ref{V1}) are contracted
with its lower (number) indices. In the one-loop diagram this
leads to the contraction of the momenta at the vertex (\ref{BigV})
with the transverse projector in the correlator $\langle vv
\rangle_{0}$ from (\ref{linesV}), which depends on the {\it same}
momentum: $P_{ij}({\bf k})k_{i}=0$.

Therefore, the quantity in the first square brackets appears in
fact $O(g^{2})$ and, like for $\gamma^{(3)}_{F}$, the coordinates
$g_{*}$ and $u_{*}$ should be substituted into it only in the
leading-order approximations (\ref{FP1}).

It remains to note that for the diagrams Nos 2, 6, 8, and 9 the
structures $A_1$ also vanish; the mechanism is the same as for the
one-loop diagram. Therefore, there is no need to calculate these
diagrams at all: their nonvanishing contributions $A_2$ are known
exactly from (\ref{Exact}) without practical calculation.

\subsection{Calculation of the scalar quantities $A_{\lowercase{i}}$}
\label{sec:A}

\begin{figure}
\caption{Diagrams of the function $\Gamma$ from
(\protect\ref{gena}) in the one-loop and two-loop approximations.}
\begin{center}
\epsfig{figure=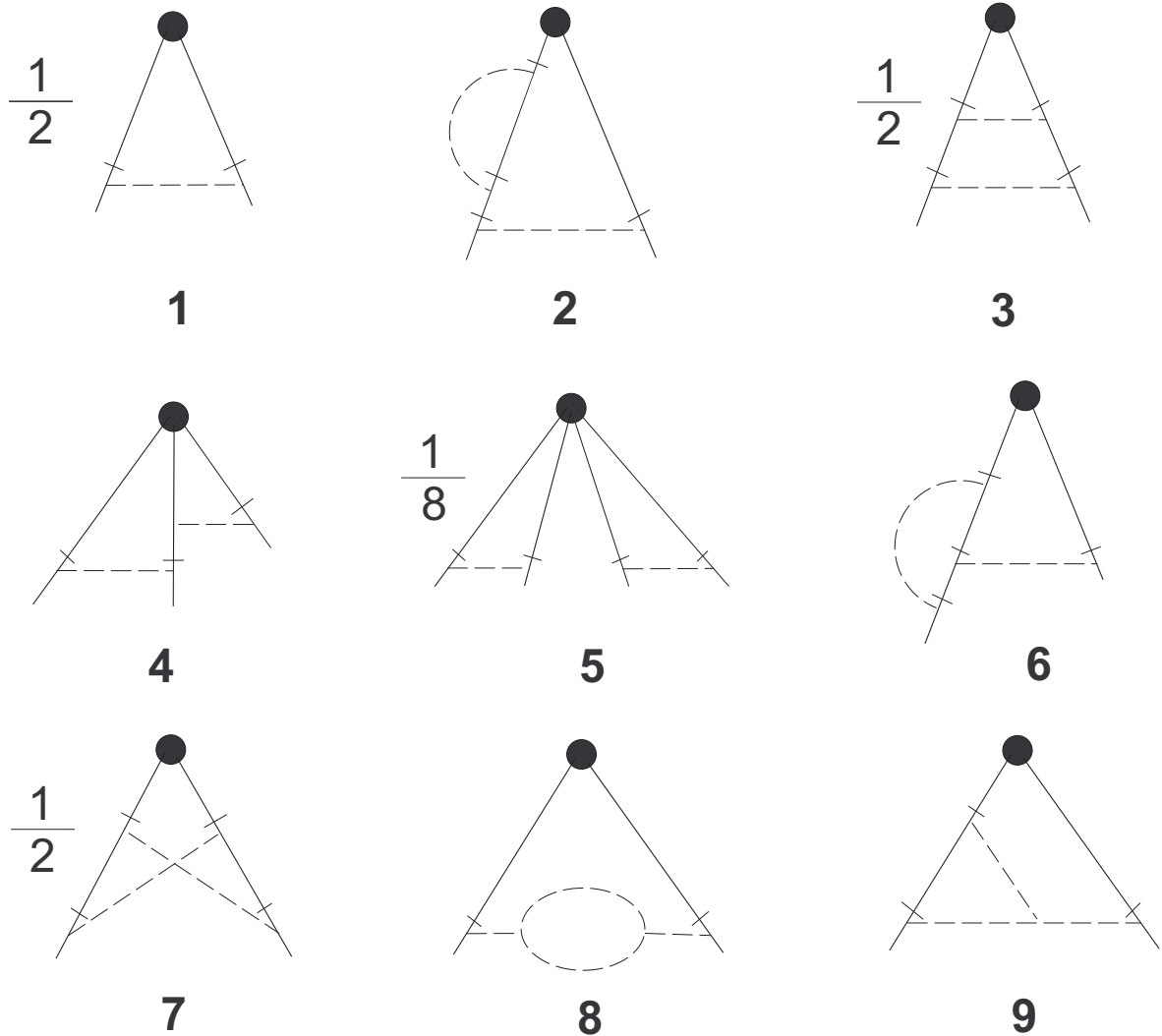, height=14cm} \label{table2}
\end{center}
\end{figure}

We shall not discuss the calculation of the scalar quantities
$A_{i}$ from Eq. (\ref{V6}) for the all diagrams from Fig.
\ref{table2} in detail, because this definitely would exceed the
readers' patience, and give only examples and general ideas. It
has much in common with the analogous calculation for Kraichnan's
model, discussed in Ref. \cite{cube} up to the three-loop level in
great detail. The present calculation differs from that of
\cite{cube} in a few respects:

(i) Diagrams Nos 8 and 9 involve the propagators $\langle v'v
\rangle_{0}$ from (\ref{linesV}) and the vertex $v'vv$ from
(\ref{vertexV}); they are absent for the case of a Gaussian
velocity field (including, of course, the case of Kraichnan's
model).

(ii) Diagrams Nos 6 and 7 are present for any Gaussian velocity
field with finite correlation time. However, for Kraichnan's model
they effectively contain closed circuits of retarded propagators
$\langle \theta'\theta \rangle_{0}$ and therefore vanish. It is
crucial here that for Kraichnan's model the velocity correlator
contains the $\delta$ function in time. In our case, the velocity
has finite correlation time and these diagrams give nonvanishing
contributions.

(iii) In Kraichnan's case, the diagram No 2 and, in general, all
diagrams with the self-energy insertions are easily taken into
account: it is sufficient just to drop them, and in the remaining
diagrams to replace the bare viscosity $\nu_0$ with its exact
analog; see Ref. \cite{RG3}. This is a consequence of the fact
that, in Kraichnan's case, the self-energy (that is, nontrivial
part of the 1-irreducible function $\langle \theta'\theta
\rangle_{\rm 1-ir}$) is exactly given by the simplest one-loop
diagram (the other contain closed circuits of retarded propagators
and vanish). That diagram does not depend on its frequency
argument and is simply proportional to $p^{2}$, its squared
momentum argument, and thus its contribution only leads to a
certain redefinition of the viscosity coefficient.

In the case at hand, the one-loop self-energy diagram is a
nontrivial function of $p$ and the calculation of such diagrams
also becomes nontrivial. In higher orders, diagrams with multiloop
self-energy insertions (absent for Kraichnan's case) will also
appear.

(iv) Diagrams Nos 3 and 4 are present and nontrivial both for our
model and for Kraichnan's case (as well as for the Gaussian model
with finite correlation time). Of course, the corresponding
analytical expressions are different in these two cases due to the
difference in the explicit forms of the correlation functions
$\langle v'v \rangle_{0}$. In particular, all their contributions
for the zero correlation time are expressed in terms of
hypergeometric functions (see Ref. \cite{RG}) while for our case
this is no longer true (see below).

(v) In Kraichnan's model, the value of $g_{*}$ was given by the
one-loop approximation exactly; in our case, the higher-order
contributions are nontrivial and should be taken into account.

As we have seen in Sec.~\ref{sec:Anal}, in the two-loop
approximation the total effect of the diagrams Nos 2, 6, 8, and 9
and of the $O(\eps^{2})$ contribution in $g_{*}$ can be found from
the exact identity (\ref{Exact}) without the practical
computation, but in the three-loop and higher orders the items
(i), (iii) and (v) become nontrivial.

Consider the calculation of the one-loop diagram, No 1 in Fig.
\ref{table2}. Using the explicit forms of the propagators and
vertices in (\ref{lines}), (\ref{vertex}) and the definition we
obtain (in renormalized variables)
\begin{eqnarray}
I^{ab}_{12} = \frac{1}{2}\, \int \frac{d\omega}{(2\pi)} \int
\frac{d{\bf k}}{(2\pi)^{d}}\, k_{a}k_{b} \,
\frac{g\mu^{2\eps}\nu^3 P_{12}({\bf k}) \, k^{4-d-2\eps}}
{(\omega^{2}+\nu^2 k^{4})}\, \frac{1}{(\omega^{2}+u^{2}\nu^2
k^{4})} =
\frac{g \mu^{2\eps}}{4u(u+1)} \, \int \frac{d{\bf
k}}{(2\pi)^{d}}\, \frac{k_{a}k_{b}}{k^{2}} P_{12}({\bf k}) \,
k^{-d-2\eps},
\label{oneloop}
\end{eqnarray}
where the factor $1/2$ in front of the integral is the symmetry
coefficient (see Fig. \ref{table2}) and the three factors in the
integrand come from the vertex of the composite operator, the
propagator $\langle vv \rangle_{0}$ of the velocity field, and the
product of two propagators $\langle \theta'\theta \rangle_{0}$,
respectively. The second equality in (\ref{oneloop}) is the result
of elementary integration over $\omega$. Obviously,
$A_1=I^{ab}_{ab}=0$, the fact already mentioned in Sec.
\ref{sec:Anal}, while for $A_2=I^{aa}_{bb}$ one obtains
\begin{eqnarray}
A_2= \frac{g\mu^{2\eps}\bar S_{d}(d-1)}{4u(u+1)} \int^{\infty}_{m}
\frac{dk}{k^{1+2\eps}} = \frac{g\bar S_{d} (\mu/m)^{2\eps}(d-1)}
{8u(u+1)\eps} \label{oneloop2}
\end{eqnarray}
with $\bar S_{d}$ from (\ref{ZZ}); the pole part of this
expression is simply obtained by the replacement
$(\mu/m)^{2\eps}\to1$. Substituting this expression into
(\ref{Atas1}) along with the expressions (\ref{FP1}) for the fixed
points gives the one-loop result (\ref{Qnp}).

Now let us turn to the calculation of the two-loop diagrams Nos
2--7 in Fig. \ref{table2}. These depend on two integration momenta
${\bf k}$ and ${\bf q}$ and two frequencies, which can always be
assigned to the two $\langle vv \rangle_{0}$ lines of a diagram.
The integrations over the frequencies (or, equivalently, the times
in the time-momentum representation) are always elementary. The
result can be interpreted as a sum of terms, each of which
corresponding to one possible ``temporal version'' of the diagram;
different temporal versions correspond to all possible orderings
of the integration times in the diagram. To each version
corresponds an ``energy denominator,'' given by the product of the
factors corresponding to all ``temporal cross-sections'' of the
diagram; to each cross-section corresponds the sum of ``energies''
${\cal E}_{\k} = \nu k^{2}$ for all intersected $\langle vv
\rangle_{0}$ or $\langle v'v \rangle_{0}$ lines and ${\cal E}_{\k}
= u \nu k^{2}$ for $\langle \theta\theta'\rangle_{0} $ lines.
Thus, with some experience, it is possible to write down the
result of the temporal integration without performing the actual
integration. As a typical example, we give the result of temporal
integration for the quantity $I^{ab}_{12}$ corresponding to the
diagram No 3:
\begin{equation}
I^{ab}_{12}= \frac{(g\mu^{2\eps})^{2}} {4u(1+u)(2\pi)^{2d}}
\int\int \frac{d{\bf k}d{\bf q}} {(kq)^{d+2\eps}} \,
\frac{(k+q)_{a}(k+q)_{b} q_{i}q_{j}P_{ij}({\bf k})P_{12}({\bf q})}
{ q^{2}({\bf k}+{\bf q})^{2} \left[k^{2}+uq^{2}+u ({\bf k}+{\bf
q})^{2} \right]} \left\{ \frac{1}{\left[k^{2}+q^{2}+u ({\bf
k}+{\bf q})^{2} \right]}+ \frac{1}{uq^{2}} \right\}, \label{DDT}
\end{equation}
now the corresponding scalar quantities $A_i$ are easily obtained.
It is convenient to represent the denominators as products of
simpler factors, and to combine the quantities $A_i$ corresponding
to different diagrams; this sometimes leads to noticeable
simplifications of the integrands. With the only exception (see
Eq. (\ref{uzhas}) below), all these quantities can be reduced to
linear combinations of the following ``basis'' scalar integrals:
\begin{equation}
\frac{1} {(2\pi)^{2d}} \int\int \frac{d{\bf k}d{\bf q}}
{(kq)^{d+2\eps}} \, \frac {({\bf k}{\bf q}) \sin^{2p}\vartheta}
{k^{2}+q^{2}+2 \beta({\bf k}{\bf q})}  \equiv \frac{\bar S_{d}^2
m^{-4\eps}}{8\eps} \Psi_{2p} (\beta),
\label{Double}
\end{equation}
where the parameter $\beta$ takes different values: $\beta=1$,
$\beta=u/(1+u)$ or $\beta=(u/2(1+u))^{1/2}$, while $p=1$, 2,
and~3, and $\vartheta$ denotes the angle between the vectors ${\bf
q}$ and ${\bf k}$, so that $({\bf k}{\bf q})= kq \cos\vartheta$.
[We do not discuss much simpler integrals, e.g. those that can be
factorized into two independent integrals over ${\bf k}$ and ${\bf
q}$, and so on.] The integrands in (\ref{Double}) involve three
independent parameters, the moduli $k$ and $q$ and the angle
$\vartheta$, so that the integrals can be written as
\[ \frac{1} {(2\pi)^{2d}}\int d{\bf k} \int  d{\bf q}\, {\cal F}(k,q,\vartheta)
= \bar S_{d}^2 \, \int dk k^{d-1} \, \int dq q^{d-1} \,
\bigl\langle {\cal F}(k,q,\vartheta) \bigr\rangle \] where the
brackets denote the angular averaging over the unit sphere in $d$
dimensions normalized such that $\langle 1 \rangle  =1$. Let us
expand the integrands in (\ref{Double}) in $\beta$ or,
equivalently, in the scalar product $({\bf k}{\bf q})= kq
\cos\vartheta$. In each term of the resulting expansion, the
integrations over the angles can be computed using the following
formulas:
\begin{equation}
\bigl\langle \cos^{2n} \vartheta
\bigr\rangle = \frac{(2n-1)!!} {d(d+2)\dots(d+2n-2)},
\label{L2}
\end{equation}
with $n=1,2,\dots$. The remaining integrals over the moduli have
the forms:
\begin{equation}
I_{n} (m) \equiv \int _{m}^{\infty}\frac {dk}{k^{1+2\eps}} \,
\int _{m}^{\infty}\frac {dq}{q^{1+2\eps}}\,
\left( \frac{kq} {k^{2}+q^{2}}\right)^{2n+2} = m^{-4\eps}\,I_{n} (1).
\label{L8}
\end{equation}
Using the identity
\begin{equation}
I_{n} (m) = - \frac{1}{2\eps} {\cal D}_{m} I_{n} (m), \qquad
{\cal D}_{m} \equiv m \partial / \partial m,
\label{L9}
\end{equation}
which follows from the last equality in Eq.~(\ref{L8}), the integral
$I_{n} (m)$ can be represented in the form
\begin{equation}
I_{n} (m) = \frac{m^{-4\eps}}{2\eps} \, \int^{\infty}_{1}
\frac {dk}{k^{1+\eps}} \, \left( \frac{k} {k^{2}+1}\right)^{2n+2},
\label{L10}
\end{equation}
that is, the number of integrations is reduced and the pole in
$\eps$ is isolated explicitly. We need only the pole part of the
integral $I_{n} (1)$, which now is simply obtained by setting
$\eps=0$ in the integrand of (\ref{L10}). The resulting integral
is easily calculated:
\begin{equation}
I_{n} (1) = \frac{1}{8\eps} \, \frac{(n!)^2}{(2n+1)!}
+O(\eps^{0}). \label{L11}
\end{equation}

Thus we have represented the pole part of the integrals (\ref{Double})
as infinite series with known coefficients. It is not difficult to see
that these series can be reduced to the hypergeometric function
\begin{equation}
F(a,b;c;z)\equiv 1+\frac{ab}{c}\, z+ \frac{a(a+1)b(b+1)}{c(c+1)}
\cdot \frac{z^{2}}{2!}+\dots,
\label{hyper}
\end{equation}
namely, for the quantities $\Psi_{2p}(\beta)$ defined in
(\ref{Double}) one obtains
\begin{equation}
\Psi_{2p}(\beta)=  \frac{- \Gamma(d/2)\Gamma(d/2-1/2+p)}
{\Gamma(d/2-1/2)\Gamma(d/2+1+p)} \,\beta\,
F(1,1;d/2+1+p;\beta^{2}), \label{PsiObsii}
\end{equation}
with Euler's $\Gamma$ function. For the first special values of
$p$ this gives:
\begin{eqnarray}
\Psi_{2} (\beta) &=& \frac {2(1-d)}{d(d+2)} \,\beta
\,F(1,1;d/2+2;\beta^{2}),
\nonumber \\
\Psi_{4} (\beta) &=& \frac{2(1-d^{2})}{d(d+2)(d+4)} \,\beta
\,F(1,1;d/2+3;\beta^{2}),
\nonumber \\
\Psi_{6} (\beta)&=& \frac{2(1-d^{2})(d+3)}{d(d+2)(d+4)(d+6)} \,
\beta \,F(1,1;d/2+4;\beta^{2}),
\label{Psis}
\end{eqnarray}
and so on. The integral
\begin{equation}
\frac{1} {(2\pi)^{2d}} \int\int \frac{d{\bf k}d{\bf q}}
{(kq)^{d+2\eps}} \, \frac {({\bf k}{\bf q})\, k^{2}\,
\sin^{4}\vartheta} {({\bf k}+{\bf q})^{2} \left[({\bf k}+{\bf
q})^{2}+q^{2}+xk^{2}\right]} \equiv \frac{m^{-4\eps} \bar
S_{d}^{2}}{2\eps}\, {\cal J}(x,d) + O(\eps^0)
\label{uzhas}
\end{equation}
(where $x\equiv 1/u$) does not reduce to the hypergeometric
function and can only be expressed in the form of a single
convergent integral, suitable for numerical calculation, for
example
\begin{eqnarray}
{\cal J} (x,d) &=& {\displaystyle \frac {(1-d^2)}{2d(d+2)(d+4)}
\int^1_0 \, dz\, \frac {1}{(1+z)(1+xz)^2}\, F \left(1,2;d/2+3;
(1+z)^{-1}(1+xz)^{-1}\right) }
 \label{Juha}
\end{eqnarray}
or
\begin{eqnarray}
{\cal J} (x,d) &=& {\displaystyle
\frac{\Gamma(d/2)}{\sqrt{\pi}\,\Gamma((d-1)/2)}\, \int^1_0 dz\,
\frac{(1-z^2)^{d/2}} {(x-1)^2+4xz^2}\, \Biggl\{ z^2 (1-z^2)^{1/2}
\ln \biggl( \frac{1+x}{2} \biggr) - z (x-1+2z^2) \arcsin z - }
\nonumber \\
&-& {\displaystyle \frac{z(1-z^2)^{1/2}(1-x-z^2)}
{[2(1+x)-z^2]^{1/2}}\, \arctan \biggl[ \frac{z [2(1+x)-z^2]^{1/2}}
{(1+x-z^2)} \biggr] \Biggr\} }.
\label{J}
\end{eqnarray}
It remains to note that in Eq. (4.3) of Ref. \cite{Juha2} the
latter integral is given with a misprint.

\section{Anomalous exponents to order $\varepsilon^{2}$}
\label{sec:Exponents}

Using the techniques described in the preceding Section, we have
performed complete two-loop calculation of the critical dimensions
$\Delta_{nl}$ of the composite operators (\ref{Fnl}) for arbitrary
values of $n$, $l$, and $d$ and obtained the following expression
for the second coefficient in expansion (\ref{answer}):
\begin{eqnarray}
\Delta_{nl}^{(2)} &=& \frac{4}{9(d-1)^{2}(d+2)^{2}(d+4)} \biggl\{ 2(d+4)
{\cal A} \left[ n(n-2)(d-1)+  \lambda_{l} \right] + (n-2)\Bigl\{6{\cal B}
\left[n(n-4)(d-1)+ 3\lambda_{l} \right]+
\nonumber \\
&+& 9{\cal C} \left[n(d+n)(d-1) - \lambda_{l}(d+1) \right]\Bigr\}\biggr\}
\label{answer2}
\end{eqnarray}
with $\lambda_{l}\equiv l(d+l-2)$ and
\begin{eqnarray}
{\cal A} &=& \frac{(x-1-1/x)(d+1)}{2(d+2)(1-x)} + \frac{(d+1)}
{2(d+4)(1-x)x(1+x)^{2}} F_{3}\left( \frac{1}{(x+1)^{2}} \right)  +
\frac{2xd(d+2)} {(1-d)(1-x)} {\cal J}(x,d),
\nonumber \\
{\cal B} &=& \frac{(d+1)}{3(1-x)^{2}(d+4)} \left[\frac{x}{x+1}
F_{3} \left(\frac{1}{2(x+1)}\right) - \frac{1}{(x+1)^{2}} F_{3}
\left(\frac{1}{(x+1)^{2}}\right) - \frac{x^{2}}{4} F_{3}
\left(\frac{1}{4}\right) \right],
\nonumber \\
{\cal C} &=& \frac{1} {9(1-x)^{2}} \biggl\{ \frac{3x^{2}(d-1)}{4}
F_{2} \left(\frac{1}{4}\right) - \frac{x[2d-1+x(d-2)]}{(x+1)} F_{2}
\left(\frac{1}{2(x+1)}\right) + \frac{[d+1+2x(d-2)]} {(x+1)^{2}} F_{2}
\left( \frac{1}{(x+1)^{2}}\right) -
\nonumber \\
&-& \frac{x^{2}(d+1)}{(d+4)} F_{3} \left(\frac{1}{4} \right) +
\frac{4x(d+1)}{(x+1)(d+4)} F_{3}\left(\frac{1}{2(x+1)}\right) -
\frac{4(d+1)} {(x+1)^{2}(d+4)}
F_{3}\left(\frac{1}{(x+1)^{2}}\right) \biggr\}.
\label{ABC}
\end{eqnarray}
Here ${\cal J}(x,d)$ is the integral (\ref{J}), $x\equiv
1/u_{*}^{(0)}$ with $u_{*}^{(0)}$ from (\ref{FP1}), and
$F_{k}(z)\equiv F(1,1;d/2+k;z)$ is the hypergeometric function
(\ref{hyper}). The values of $F_{k}$ entering into (\ref{ABC}) can
be related by the recurrent relation
\[(z-1)F_{2}(z)= z(d+2)F_{3}(z)/(d+4)-1,\]
but the resulting expressions look more cumbersome and we have
kept both $F_{2}$ and $F_{3}$ in the formulas.

Contributions with ${\cal A}$, ${\cal B}$, and ${\cal C}$ in
(\ref{answer2}) come from the structures $A_{1}$ with $k=2$,
$A_{1}$ with $k=3$, and $A_{2}$ with $k=3$, respectively. The
structure $A_{2}$ with $k=2$ gives no contribution to
$\Delta_{nl}^{(2)}$, as discussed in Sec.~\ref{sec:Anal} in
connection with Eq.~(\ref{Exact}). For the most interesting case
$d=3$ one obtains:
\begin{equation}
{\cal A}= -0.90239, \qquad {\cal B}= -0.135498, \qquad {\cal C}=
0.19622, \qquad {\cal J}= -0.024976. \qquad
 \label{ABCJ}
\end{equation}

Expression (\ref{answer2}) simplifies for the most important case of the
isotropic sector (even $n$ and $l=0$):
\begin{eqnarray}
\Delta_{n0}^{(2)} =  \frac{n(n-2)}{(d-1)(d+2)^{2}(d+4)} \biggl\{
2(d+4) {\cal A}+ 6(n-4) {\cal B}+9(d+n) {\cal C} \biggr\}.
\label{L=0}
\end{eqnarray}

For the simplest nontrivial case $n=4$ one obtains
\begin{eqnarray}
\Delta_{40}^{(2)} = 8(2{\cal A}+9{\cal C})/(d-1)(d+2)^{2},
\label{L=0,N=4}
\end{eqnarray}
that is, the quantity ${\cal B}$ does not enter into the result. For
$n\ge6$, all the coefficients (\ref{ABC}) contribute to the result.

\section{Discussion and conclusion} \label{sec:Conclusion}

We have studied a model of a passive scalar field, governed by the
diffusion-advection equation (\ref{1}) and subject to a
large-scale random forcing (\ref{2}). The advecting velocity field
obeys the Galilean-invariant Navier-Stokes equation (\ref{1.1})
subject to an external random force, white in time and having a
power-law spectrum $\propto k^{4-d-2\eps}$; see Eqs. (\ref{1.2})
and (\ref{1.9}).

Using the RG and OPE methods, we have shown that the structure
functions of the scalar field display anomalous scaling behavior;
see Eqs. (\ref{Anomal}) and (\ref{Animal}). The corresponding
anomalous exponents $\Delta_{n}$ are identified with the critical
(scaling) dimensions of certain composite fields (operators),
namely, powers of the local dissipation rate of scalar
fluctuations (\ref{Fn}), which offers the possibility to calculate
them within a regular perturbation theory, as series in $\eps$;
see Eq.~(\ref{answer}).

The calculation has been accomplished to the second order,
$\eps^{2}$ (two-loop approximation), including the exponents of
anisotropic contributions (\ref{Animal}). The latter are
identified with the critical dimensions of tensor composite fields
built of the scalar gradients (\ref{Fnl}). The first-order
expressions (\ref{Qnp}) coincide with the exponents of the
well-known Kraichnan's rapid-change model (up to a simple
normalization), while in the second order they are different. Like
for the rapid-change model, the second-order structure function is
not anomalous.

Thus we have overcome two important limitations of the previous
treatments of the problem: absence of time correlations and
Gaussianity of the advecting velocity field. It is interesting to
note that both the RG-mechanism of the anomalous scaling and the
results for the exponents are, in many respects, similar to the
case of the rapid-change model. Let us compare our findings with
those for the Gaussian models.

{\bf Universality: Independence of the forcing and relevance of
the zero-modes picture.} As we have seen, the critical dimensions
of all composite operators (\ref{Fn}) and (\ref{Fnl}), and
therefore the corresponding anomalous exponents (including
anisotropic sectors), are independent of the forcing, specified by
the correlator (\ref{2}). In particular, this means that they
remain unchanged if the stirring noise in Eq. (\ref{1}) is
replaced by an imposed constant gradient, like e.g. in Refs.
\cite{LM,RG3,OU}. The role of the forcing is to maintain the
steady state of the system and thus to provide nonvanishing
amplitudes for the power-law terms with those universal exponents.
This behavior is already well known for the passive scalar or
vector fields, advected by the Gaussian velocity fields with
vanishing or finite correlation time.

In the language of the RG (which is equally applicable to the case
of a zero or finite correlation time of the advecting field) this
is explained as follows: the stirring force or the imposed
gradient do not enter into the diagrams that determine the
renormalization of the operators (\ref{Fn}) and (\ref{Fnl}), so
that their dimensions appear forcing-independent. Similar diagrams
determine the contributions of those operators into the
operator-product expansions (\ref{OPE}), which are nontrivial even
for the unforced model. The difference is that for the unforced
model, the mean values of the operators vanish, so that they give
no contribution to the right-hand sides of representations like
(\ref{OR}). For the isotropic correlator (\ref{2}), scalar
operators acquire nonzero mean values and contribute to the
right-hand side of (\ref{OR}), while for the anisotropic
correlator or the imposed constant gradient, the mean values of
irreducible tensor operators also become nonzero and their
contributions are ``activated'' in representations (\ref{OR}).

For the case of a Gaussian advecting field with vanishing
correlation time, when the equal-time correlations functions
satisfy exact closed differential equations, the above picture it
is easily understood in the language of the zero-mode approach
\cite{FGV}: forcing terms do not affect the corresponding
differential operators; thus the anomalous exponents, determined
by the zero modes (solutions of homogeneous unforced equations)
also appear forcing-independent. On the contrary, the amplitudes
are determined by the matching of the inertial-range zero-mode
solutions with the forced large-scale solutions, which is only
possible in the presence of the forcing terms.

The exact resemblance in the RG picture of the rapid-change models
and the finite-correlated cases suggests that for the latter, the
concept of zero modes (and thus of statistical conservation laws)
is also applicable, although the corresponding equations are not
differential and involve infinite diagrammatic series.

{\bf Hierarchy of anisotropic contributions.} In the presence of
large-scale anisotropy (that is, the anisotropy introduced at
scales of order $L$ by the forcing in Eq. (\ref{1})), structure
functions of the scalar field can be decomposed in irreducible
representations of the $d$-dimensional rotation group SO$(d)$.
Such a decomposition naturally arises from the corresponding OPE,
provided it is made in irreducible traceless tensor composite
operators; the rank $l$ of a tensor operator can be used to label
the terms of the SO$(d)$-expansion and can be viewed as the
measure of anisotropy of the corresponding term (``sector''). Thus
each anisotropic sector is characterized by its own set of scaling
exponents, the leading term is given by the $l$-th rank composite
operator with minimal critical dimension.

Explicit expressions for these dimensions, derived to second order
in $\eps$, exhibit an hierarchy related to the degree of
anisotropy: the higher is the rank of the operator (the more
anisotropic is the contribution), the larger is the corresponding
dimension, and thus the less important is its contribution to the
inertial-range behavior. This hierarchy can be expressed by the
relation $\partial \Delta_{nl} / \partial l >0$, which is obvious
from the first-order expression (\ref{Qnp}). It holds for all
values of $n$ and $d$. This picture is similar to the hierarchy
relations derived earlier for the passive scalar and magnetic
fields advected by the Gaussian velocity ensembles
\cite{LM,RG3,Juha2,RG4,RG5}.

In particular, this means that the overall leading term is given
by the exponent from the isotropic sector, and it is therefore the
same for the isotropic and anisotropic forcing. It also should be
stressed that the independence of the scaling behavior in
different sectors is a direct consequence of the linearity of our
model, independence of the exponents on the random force, and the
SO$(d)$ symmetry of the unforced model. On the contrary, the {\it
hierarchy} of the exponents follows from the explicit expressions,
obtained only by practical calculation.

According to the Kolmogorov--Obukhov theory \cite{Monin,Legacy},
the anisotropy introduced at large scales by the forcing (boundary
conditions, geometry of an obstacle {\it etc}) dies out when the
energy is transferred down to smaller scales owing to the cascade
mechanism (isotropization of the developed turbulence in the
inertial-range). The analytical results discussed above confirm
this classical concept and give a more quantitative picture of the
isotropization.

The hierarchical picture, derived here and in Refs.
\cite{LM,RG3,Juha2,RG4,RG5} for passively advected fields, appears
unexpectedly general, being compatible with that established
recently in the field of NS turbulence, on the basis of numerical
simulations and natural experiments; see Refs. \cite{Arad1} and
references therein. There, the velocity structure functions were
decomposed in the irreducible representations of the rotation
group. It was shown that in each sector of the decomposition,
scaling behavior can be found with apparently universal exponents.
The amplitudes of the various contributions are nonuniversal,
through the dependence on the position in the flow, the local
degree of anisotropy and inhomogeneity, and so on.

It is worth recalling here that the so-called ``additive fusion
rules,'' hypothesized for the NS turbulence in a number of papers
(see e.g. Ref. \cite{Eyink2}) and characteristic of the models
with multifractal behavior (see Ref. \cite{DL}), arise naturally
in the context of the rapid-change models owing to their {\it
linearity}. The existing results for the Burgers turbulence can
also be interpreted naturally as a consequence of similar fusion
rules, where only finite number of dangerous operators contributes
to each structure function; see Ref. \cite{Burg1}. This is rather
surprising because the equations for the correlation functions in
such cases are neither closed nor isotropic and homogeneous. One
can thus speculate that the anomalous scaling for the genuine
turbulence can also appear, in some sense, a linear phenomenon. Of
course, one should not insist too much on this bold assumption.

{\bf Universality: Independence of the time scales.} An important
issue is that of the universality of anomalous exponents. As
already discussed, the exponents $\Delta_{nl}=\Delta_{nl}(\eps,d)$
in Eq. (\ref{answer}) are independent of the forcing in the scalar
equation (\ref{1}), and thus independent of all the parameters
that can appear in its correlation function (\ref{2}).

However, the exponents depend on the exponent $\eps$ that enters the
correlation function of the stirring force (\ref{1.2}) in the NS equation
(\ref{1.1}). They also depend on $d$, the dimensionality of the ${\bf x}$
space (note that the basis dimensions related to the velocity field are
$d$-independent; see Eqs. (\ref{Deltas})).

Earlier, it was argued on phenomenological grounds that the
anomalous exponents of the scalar field can depend on more
characteristics of the advecting field than only the exponents;
see e.g. the discussion in Ref.~\cite{ShS}. Indeed, analytical
derivation of the anomalous exponents of the passive scalar field,
advected by a Gaussian velocity with finite correlation time, has
revealed for some asymptotic regimes (``local turnover exponent'')
their dependence on the correlation time of the velocity field
(more precisely, the dimensionless ratio of the correlation times
of the scalar and velocity fields); see
Refs.~\cite{RG3,Juha2,Falk3}.

In our case, the exponents could depend, in principle, on the
analogous dimensionless parameter $u_{0}\equiv \kappa_{0}/\nu_0$
from (\ref{Prad}), the (inverse) Prandtl number. After the RG
resummation, this parameter is replaced with the corresponding
invariant variable, which has exactly the meaning of the ratio of
the scalar and velocity correlation times (for a detailed
discussion of this point see Ref. \cite{RG3}). However, the
analysis of the RG equation shows that in the IR asymptotic range,
this parameter tends to a fixed point, whose coordinate $u_{*}$
depends on $d$ and $\eps$, but not on the initial value $u_{0}$;
see Eqs. (\ref{FPE}), (\ref{FP1}). As a result, all the
dimensions, including $\Delta_{nl}$ from (\ref{answer}), appear
independent of $u_{0}$. In the RG language, the nonuniversality
(that is, the dependence on the ratio $u_{0}$ or its analog) of
the exponents in the Gaussian model is a consequence of the
infinite degeneracy of the IR stable fixed point; see the
discussion in \cite{RG3}. In the NS model, the fixed point is
unique (non-degenerate), and the exponents appear universal.

We stress that, although the coordinates of the fixed point are
known only in the two-loop approximation (see the discussion below
Eqs. (\ref{FPE}), (\ref{FP1})), the statement about the
universality is exact, that is, it holds to all orders of the
$\eps$ expansion.

Since the degeneracy of the fixed point in the model studied in
Refs.~\cite{RG3,Juha2,Falk3} is an artifact of the Gaussianity of
the velocity ensemble, we believe that our result for the
non-Gaussian velocity ensemble, described by the
Galilean-covariant NS equation, suggests that for the real passive
advection the anomalous exponents are universal, that is,
independent of the Prandtl number or the ratio of the scalar and
lvelocity correlation times. This is probably the most important
qualitative conclusion that can be inferred from our analysis. It
is then relevant to discuss the role played by the Galilean
symmetry of our model in the RG-analysis.

{\bf Sweeping effects and the Galilean invariance.} The results
obtained within the RG and OPE approach and within the $\eps$
expansion, are reliable and internally consistent for
asymptotically small $\eps$. A serious question is that of the
validity of the $\eps$ expansion for finite $\eps$'s, and the
possibility of the extrapolation of those results to the physical
value $\eps=2$.

For the rapid-change model, the $\eps$ expansion works
surprisingly well. It was demonstrated \cite{cube} that the
knowledge of three terms allows one to obtain reasonable
predictions for finite $\eps\sim1$; even the plain $\eps$
expansion captures some subtle qualitative features of the
anomalous exponents established in the exact solutions of the
zero-mode equations and in numerical simulations. The quantitative
agreement can be achieved with the aid of various improvements,
like the inverse $\eps$ expansion or interpolation formulas
\cite{cube}.

In the case of the Gaussian model with a finite correlation time,
however, there is a natural upper bound for the range of validity
of the $\eps$ expansion: for $\eps$ larger than certain threshold
value $\eps_{c}$, the velocity field (and hence all its powers)
become dangerous: their critical dimensions, known exactly due to
the Gaussianity, become negative, and new strong IR singularities
occur in the diagrams; see the discussion in Ref. \cite{RG3}. This
leads to a qualitative changeover in the small-$mr$ behavior of
the scalar field, as demonstrated in Refs. \cite{AM,AM2,Horvai}
using certain nonperturbative analytical methods and numerical
experiments. Therefore, the results obtained within the plain
$\eps$ expansion no longer apply.

Physically, this is a manifestation of the fact that above $\eps_{c}$, the
so-called sweeping effects (kinematic transfer of the small-scale turbulent
eddies by the large-scale ones) become important. Thus such threshold value
can also be viewed as the upper bound of the range of validity of the model
itself: the lack of Galilean covariance becomes a serious shortcoming of the
synthetic Gaussian velocity ensemble when the sweeping effects become
important.

In the model (\ref{1.1}), (\ref{1.2}), the dimensions of the
powers $v^{n}$ are known exactly, $\Delta[v^{n}]=n \Delta[v]
=n(1-2\eps/3)$; they all become negative for $\eps>\eps_{c}=3/2$
\cite{JETP,UFN,turbo}. Some operators, built of the velocity field
and its temporal derivatives, also become dangerous for $\eps<2$;
see \cite{UFN,turbo}. Their contributions to the OPE for the
correlation functions of the velocity and scalar fields become
singular; however, they can be summed out explicitly using certain
infrared perturbation theory. This indeed results in a qualitative
changeover in the small-$mr$ behavior of the correlation
functions, their strong dependence on the IR scale $\ell=1/m$, and
superexponential decay in time \cite{JETP,UFN,turbo}, in agreement
with the phenomenological analysis of Refs.~\cite{CK}.

Galilean symmetry of our model guarantees, however, that the
invariant quantities, for example, the equal-time structure
functions (\ref{struc}), are not affected by the sweeping. More
formally, the contributions of the aforementioned dangerous
operators do not appear in the OPE for Galilean invariant
correlation functions; see Refs. \cite{Book3,JETP,UFN,turbo,Komp}
for detailed discussion. This means that in model
(\ref{1})--(\ref{1.9}), the scaling relations obtained for small
$\eps$, for Galilean invariant quantities can be extrapolated
beyond the threshold $\eps_{c}$, in spite of the fact that the
sweeping becomes important there. The most recent numerical
simulations of the model (\ref{1.1})--(\ref{1.9}) have shown that
the scaling relations, obtained by the RG analysis for the
structure functions, remain valid for $\eps$ as high as $\eps=7/4$
\cite{Biferale}.

{\bf Extrapolation to the physical value $\eps=2$. Relevance of
the model for the real turbulent advection.} Our calculation of
the anomalous exponents implied smallness of the RG expansion
parameter $\eps$. For small $\eps$, a serious flaw shared by our
model with the Gaussian ones is that the advecting velocity field
is non-intermittent, in contrast to the real turbulent fluid.
However, numerical simulations of Refs. \cite{Biferale,Pandit}
suggest that, as $\eps$ increases, the behavior of the model
(\ref{1.1})--(\ref{1.9}) undergoes a qualitative changeover and
the scaling of the velocity field becomes anomalous: the exponents
of the structure functions become different from the results of
naive extrapolation of the small-$\eps$ prediction. In the RG
language this probably means that certain Galilean invariant
operators acquire negative critical dimensions for some finite
values of $\eps$, close to the physical value $\eps=2$.
Unfortunately, identification of those operators and calculation
of their dimensions on the basis of the model
(\ref{1.1})--(\ref{1.9}) lies beyond the scope of the present RG
technique: the effect takes place for finite, and not small,
values of $\eps$, while the dimensions of the operators are known
only in the form of the first terms of the expansions in $\eps$
(some dimensions are known exactly, but they all remain positive
for $\eps\le2$). Detailed discussion of the critical dimensions of
Galilean invariant operators can be found in Refs.
\cite{Book3,UFN,turbo} and the original papers \cite{Pismak,Kim}.
Hopefully, the problem will be solved with the aid of an
alternative perturbation theory (the expansion in $1/d$ seems very
promising, but so far it has been constructed only for Kraichnan's
model and only to the leading order \cite{Falk1}).

If the dangerous Galilean invariant operators indeed arise in the
model (\ref{1.1})--(\ref{1.9}) for some finite values of $\eps$,
they will also contribute to the OPE's for the structure functions
(\ref{struc}) of the scalar field. Physically, this corresponds to
the contribution of the velocity to the intermittency of the
scalar field, while the contributions of the operators (\ref{Fn}),
(\ref{Fnl}) correspond to the intrinsic intermittency of the
scalar field itself. Obviously, only the latter contribution can
be described within the $\eps$ expansion. Since the scalar fields
appear much more intermittent than the velocity field, one can
assume that this latter contribution dominates the anomalous
behavior of the scalar, or, at least, it is relatively more
important than the former one. One can therefore hope that the
dimensions $\Delta_n$ and $\Delta_{nl}$, taken at the physical
value $\eps=2$, can be identified with the leading anomalous
exponents of the structure functions of the real passive scalar
field.

Experimental results for the structure functions of a passive
scalar field are presented in Refs. \cite{War,Ant,Helium} in terms
of the exponents $\zeta_{n}$, with $S_{n} \propto
r^{\,\zeta_{n}}$. For an even function, in our notation
$\zeta_{2n}= -2n\Delta_{\theta}+ \Delta_{n}$, where
$\Delta_{\theta}=-1+\eps/3$ from (\ref{Deltas}) is the critical
dimension of the scalar field and $\Delta_{n}$ are the dimensions
of the operators (\ref{Fn}), with the second-order expression
given in Eq. (\ref{L=0}). For the physical value $\eps=2$ this
gives $\zeta_{2n}= 2n/3+ \Delta_{n}(\eps=2)$.

The results of \cite{War,Ant,Helium} seem to be consistent with
the Kolmogorov value $\zeta_{2}= 2/3$ for the second-order
function, in agreement with our exact result $\Delta_{1}=0$.
Possible deviation, if any, can be attributed to the anisotropy of
the experimental setup or/and contribution of velocity's
intermittency, neglected in our analysis.

From the Fig.~3 presented in the most recent study \cite{Helium}
one can infer $\Delta_{2} \simeq -0.23$ and $\Delta_{3} \simeq
-0.67$, which shows clear deviation from the Kolmogorov values
($\Delta_{n}=0$ for all $n$). For $\eps=2$ and $d=3$ the one-loop
approximation (\ref{Qnp}) gives $\Delta_{2} \simeq -1$ and
$\Delta_{3} \simeq -3$; the two-loop correction (\ref{L=0})
appears numerically very small and does not affect this result
markedly. Admittedly, it is difficult to speak about a good
quantitative agreement with the experimental values.

The situation resembles that encountered for Kraichnan's model.
There, the one-loop results for the most realistic value $\xi=4/3$
are also equal to $\Delta_{2} \simeq -1$ and $\Delta_{3} \simeq
-3$ (see the remark and footnote below Eq. (\ref{Qnp})) and
essentially overestimate the real value of the exponents
$\Delta_{2} \simeq -0.3$ and $\Delta_{3} \simeq -0.7$, known from
the numerical simulations of Refs. \cite{VMF}. The second-order
corrections (although different from their analogs in our model)
also appear too small to improve the agreement. A better agreement
is achieved if the third-order correction (which is not small) is
taken into account, and the straightforward expansion in $\xi$ is
augmented by additional considerations about its nature,
convergence properties, character and location of singularities
etc; see Ref. \cite{cube}.

In contrast to Kraichnan's model, there is no reason to believe
that in our case the series in $\eps$ have finite radii of
convergence. As in most field theoretic models, they can only be
asymptotical series. Thus, one should not have expected that the
straightforward summation of the first two (or even more) terms
would give a good result. In models of critical behavior accurate
theoretical predictions for the exponents imply knowledge of the
large-order asymptotic behavior of the coefficients of the $\eps$
expansions, obtained using the instanton calculus, and special
resummation procedures for the divergent series \cite{Zinn}. In
dynamical models the corresponding methods are at their infancy
\cite{instanton,instanton2,Komarova}: to the best of our
knowledge, an instanton-type solution for an action functional of
the Martin-Siggia-Rose type has been obtained only for a model
whose equal-time correlation functions correspond to a system in
thermodynamic equlibrium \cite{Komarova}. The instanton analysis
of Refs. \cite{instanton2}, performed in Lagrangian variables
(which implies zero correlation time of the velocity field, that
is, only Kraichnan's case) did not touch upon the problem of the
large-order coefficients of perturbative series; it has mostly
been concentrated on the behavior of the exponents $\zeta_n$ at
large $n$. One can hope that further development of the instanton
techniques for dynamical models, combined with the RG framework
will give the solution of this important problem. This work is
left for the future.

\acknowledgments The authors thank Luca Biferale, Antonio Celani,
Michal Hnatich, Antti Kupiainen, Alessandra Lanotte, Andrea
Mazzino, and Paolo Muratore Ginanneschi for discussions. The work
was supported by the Nordic Grant for Network Cooperation with the
Baltic Countries and Northwest Russia No.~FIN-20/2003. N.V.A. was
also supported by the Academy of Finland (Grant No.~203122).
L.Ts.A., N.V.A., and T.L.K. thank the Department of Physical
Sciences in the University of Helsinki for their kind hospitality.
N.V.A. and J.H. thank the Organizers of the Tenth European
Turbulence Conference ``Advances in Turbulence X'' (The Norwegian
University of Science and Technology, Trondheim, Norway, 29 June
--2 August 2004).

\begin{table}
\caption{Canonical dimensions of the fields and parameters in the
model (\protect\ref{action}), (\protect\ref{actionV}).}
\label{table1}
\begin{tabular}{ccccccccc}
$F$ & ${\bf v}'$ &  ${\bf v}$ & $\theta'$ & $\theta$ & $\nu$, $\nu _{0}$ &
$m$, $\mu$, $1/L$  & $g_{0}$ &  $g$, $u_{0}$, $u$
\\
\tableline
$d_{F}^{k}$ & $d+1$ & $-1$ & $d$ & $0$ & $-2$ & 1& $2\eps$ & 0  \\
$d_{F}^{\omega }$ & $-1$ & 1 & $1/2$ & $-1/2$ & 1 & 0 & 0 & 0 \\
$d_{F}$ & $d-1$ & 1 & $d+1$ & $-1$ & 0 &1&   $2\eps $ & 0 \\
\end{tabular}
\end{table}

\end{document}